\documentclass[aps,longbibliograph,longbibliography, footinbib,superscriptaddress, twocolumn,nofootinbib]{revtex4-2}
\usepackage{graphicx}
\usepackage{stmaryrd}
\usepackage{amsmath, amsfonts, mathrsfs, braket,latexsym}
\usepackage[dvipsnames]{xcolor}
\usepackage{xspace}
\usepackage{comment}
\usepackage{bm}
\definecolor{lightblue}{rgb}{0.13, 0.26, 0.99}

\usepackage{multirow}

\usepackage[
colorlinks=true,
urlcolor=blue,
citecolor=blue,
linkcolor=blue,
hyperfootnotes=false]{hyperref}

\allowdisplaybreaks

\begin{document}

\title{Electric-field control of magnetic anisotropies: \\
applications to Kitaev spin liquids and topological spin textures}
\author{Shunsuke C. Furuya}
\affiliation{Department of Physics, Ibaraki University, Mito, Ibaraki 310-8512, Japan}
\affiliation{Department of Basic Science, University of Tokyo, Meguro, Tokyo 153-8902, Japan}
\affiliation{Department of Liberal Arts, Saitama Medical University, Moroyama, Saitama 350-0495, Japan}
\affiliation{Institute for Solid State Physics, The University of Tokyo, Kashiwa, Japan}
\author{Masahiro Sato}
\affiliation{Department of Physics, Ibaraki University, Mito, Ibaraki 310-8512, Japan}
\affiliation{Department of Physics, Chiba University, Chiba 263-8522, Japan}
\date{\today}

\begin{abstract}
Magnetic anisotropies often originate from the spin-orbit coupling and determine magnetic ordering patterns.
We develop a microscopic theory for DC electric-field controls of magnetic anisotropies in magnetic Mott insulators and discuss its applications to Kitaev materials and topological spin textures.
Throughout this paper, we take a microscopic approach based on Hubbard-like lattice models, tight-binding models with on-site interactions.
We derive a low-energy spin Hamiltonian from a fourth-order perturbation expansion of the Hubbard-like model.
We show in the presence of a strong intra-atomic spin-orbit coupling that DC electric fields add non-Kitaev interactions such as a Dzyaloshinskii-Moriya interaction and an off-diagonal $\Gamma'$ interaction to the Kitaev-Heisenberg model and can induce a topological quantum phase transition between Majorana Chern insulating phases.
We also investigate the inter-atomic Rashba spin-orbit coupling and its effects on topological spin textures.
DC electric fields turn out to create and annihilate magnetic skyrmions, hedgehogs, and chiral solitons.
We propose several methods of creating topological spin textures with external electromagnetic fields.
Our theory clarifies that the strong but feasible electric field can control Kitaev spin liquids and topological spin textures.
\end{abstract}

\maketitle

\section{Introduction}\label{sec:introduction}

The spin-orbit coupling (SOC) is of significance to topological electronic states of matters~\cite{KaneMele_2005a,KaneMele_2005b,Nayak_TQC_review_2008,hasan_ti_review,qi_TI_TSC_review_2011,ando_ti_review}.
Recently, magnetic anisotropies, a direct descendent of SOC in quantum spin systems, has enjoyed renewed theoretical and experimental interests for their essential roles in topological states of magnetic materials such as topological spin textures~\cite{nagaosa_skyrmion_rev_2013,fujishiro_skyrmion_rev_2020} and Kitaev spin liquids~\cite{kitaev,trebst_kitaev_review_2017,Hermanns_kitaev_review_2018,Motome_kitaev_review_2020,Takagi_kitaev_review_2019,Janssen_kitaev_review_2019,Motome2020_JPC,motome2023_vdW}.

In general, the SOC violates a part of the symmetry that the system would \textit{a priori} possess.
Should the SOC be absent, the spin rotation symmetry would be independent of the spatial rotation symmetry.
However, since the SOC is always present in real materials, the spatial symmetry cannot be unrelated to the spin symmetry no matter how small the SOC is in that material.
Note that the SOC is a source of the magnetic anisotropy. Here, we call the magnetic anisotropy as terms in the spin Hamiltonian that break the spin rotational symmetry~\footnote{The Dzyaloshinskii-Moriya interaction and the Kitaev interaction dealt with in this paper are thus deemed magnetic anisotropies in this sense.}. 

\begin{figure}[t!]
	\centering
	\includegraphics[bb = 0 0 1000 750, width=\linewidth]{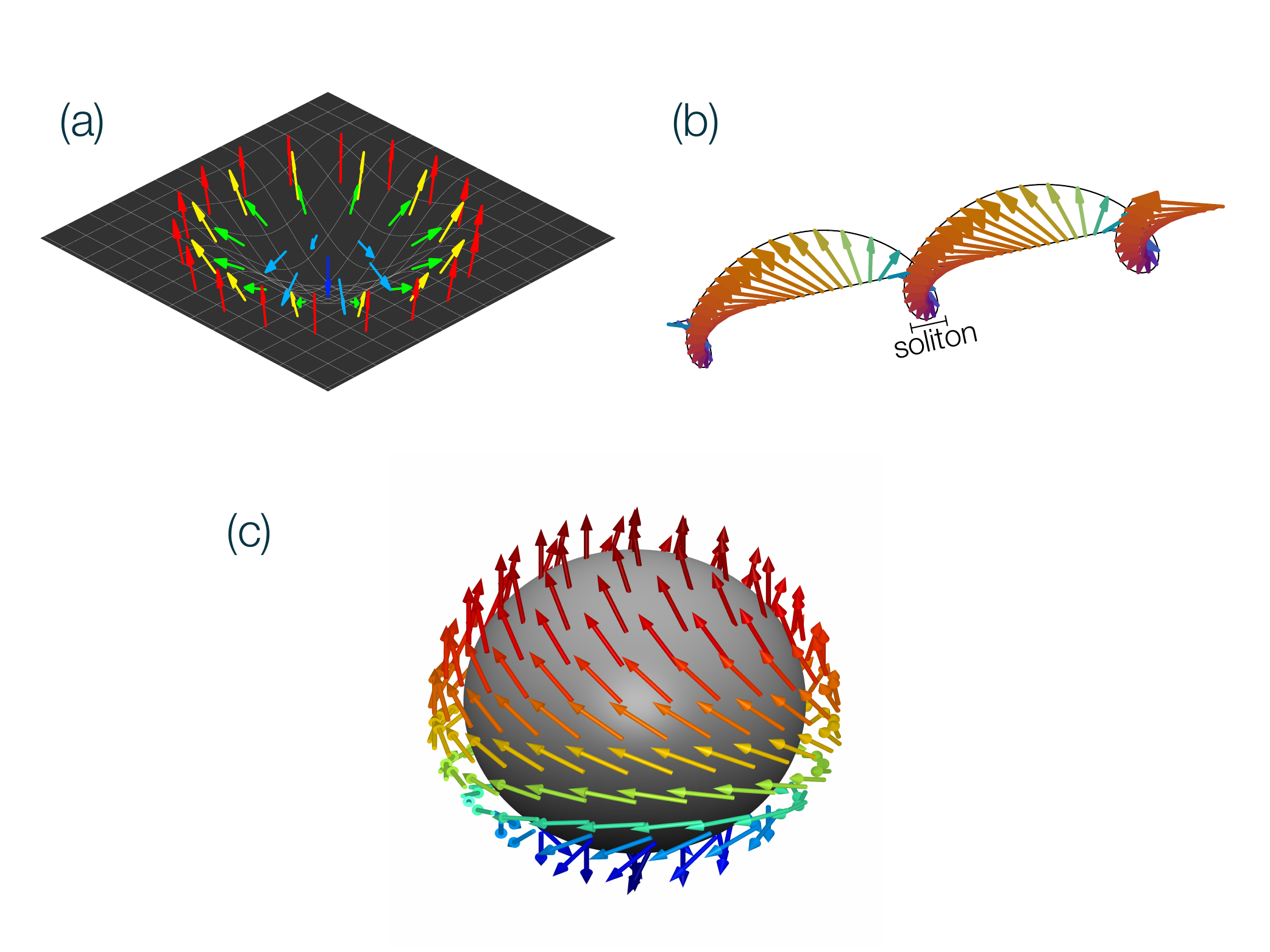}
	\caption{
    Representative topological spin textures.
    (a) N\'eel-type magnetic skyrmion.
    (b) Chiral soliton lattice (CSL) in spin chain. 
    (c) Magnetic hedgehog.
    }
	\label{fig:topological_spin_textures}
\end{figure}

An important symmetry that the SOC potentially breaks is a spatial inversion symmetry.
Violating the inversion symmetry, the SOC gives rise to an antisymmetric magnetic anisotropy known as the Dzyaloshinskii-Moriya interaction (DMI)~\cite{dzyaloshinskii,moriya}. 
The competition between the DMI and ferromagnetic exchange interactions yields topological spin textures such as magnetic skyrmions [Fig.~\ref{fig:topological_spin_textures}~(a)]~\cite{Muhlbauer2009,Yu2010,Seki2012,Kzsmrki2015,Kurumaji2017}, chiral solitons [Fig.~\ref{fig:topological_spin_textures}~(b)]~\cite{togawa_csl,Kishine2015_csl,Togawa2016_csl}, and magnetic hedgehog [Fig.~\ref{fig:topological_spin_textures}~(c)]~\cite{kanazawa2011_hedgehog,Tanigaki2015_hedgehog,Kanazawa2016_hedgehog,Fujishiro2019_hedgehog}.
The topological spin texture carries a nonzero topological index.
The nonzero topological index makes the topological spin textures robust against disturbances and prevent them from decaying into topologically trivial spin patterns such as the ferromagnetically ordered state.
The topological spin texture is thus promising for device applications~\cite{Nagaosa_emergent_EM_2012,nagaosa_skyrmion_rev_2013,fujishiro_skyrmion_rev_2020}.

\begin{figure}[t!]
	\centering
	\includegraphics[bb = 0 0 1000 600, width=\linewidth]{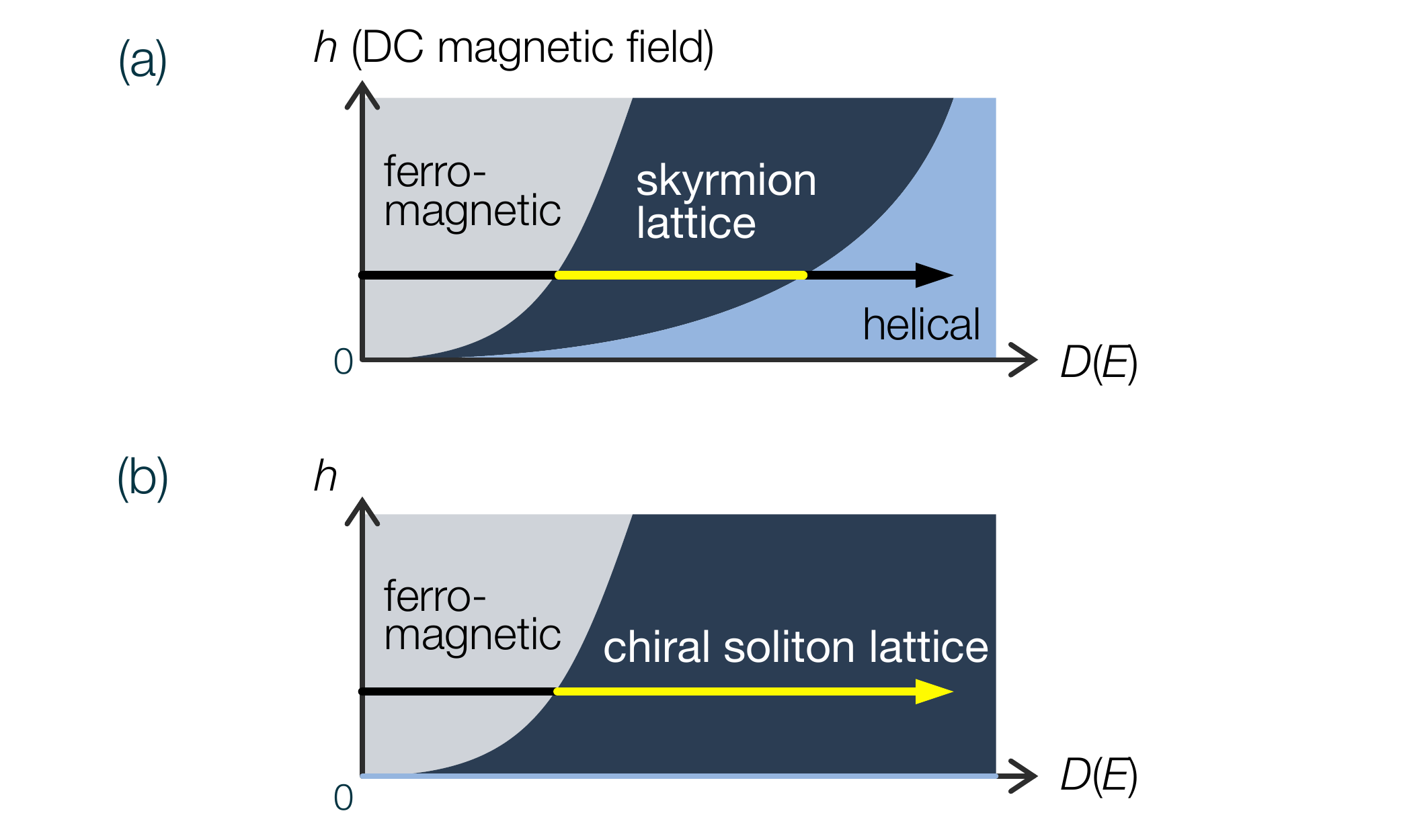}
	\caption{(a) Schematic ground-state phase diagram of square-lattice classical Heisenberg ferromagnetic model with DMI [see Eq.~\eqref{H_skx_neel}], including  helical (HL), Skyrmion-crystal (SkX), and ferromagnetic (FM) phases~\cite{Muhlbauer2009,Yu2010,Seki2012,mochizuki_skx_2012,seki2016skyrmions}.
    The vertical axis denotes the DC magnetic field $h$ and the horizontal axis denotes  the strength of the DMI, $D(\bm E)$, as a function of the DC electric field $\bm E$.
    (b) Schematic ground-state phase diagram of classical Heisenberg ferromagnetic chain with DMI [see Eq.~\eqref{H_CSL}], including CSL and FM phases~\cite{togawa_csl,Kishine2015_csl,Togawa2016_csl}. 
     }
	\label{fig:e_vs_h}
\end{figure}

The SOC can also give rise to inversion-symmetric magnetic anisotropies such as Ising interactions (e.g., $S_i^zS_j^z$), where $S_j^z$ denotes the $z$ component of a localized spin $\bm{S}_j$ at the site $j$.
This symmetric magnetic anisotropy renders the Kitaev material intriguing~\cite{Kitaev_2006,Baskaran_kitaev_2007,jackeli_kitaev_2009,knolle_kitaev_2014,yamaji_kitaev_iridates_2014,nasu_kitaev_thermal_2015,o'brien_kitaev_2016,trebst_kitaev_review_2017,Hermanns_kitaev_review_2018,bolens2018_kitaev,Takagi_kitaev_review_2019,Janssen_kitaev_review_2019,Motome_kitaev_review_2020}.
The Kitaev material, the Kitaev model and its derivatives, has drawn intensive attraction quite some time for their significance to fundamental physics and relevance for quantum computation~\cite{Kitaev_2006,Baskaran_kitaev_2007,jackeli_kitaev_2009,knolle_kitaev_2014,yamaji_kitaev_iridates_2014,nasu_kitaev_thermal_2015,o'brien_kitaev_2016,trebst_kitaev_review_2017,Motome_kitaev_review_2020}.
In light of scientific interests and engineering applications, controlling methods of these SOC-driven topological states of magnetic materials are currently one of the central issues in condensed-matter physics, quantum physics, and applied physics~\cite{Romming2013,Hsu2016,Matsuno2016_DM,Vishkayi2020_strain_electric_control,chen2021_skyrmion,kanega2021_hhg,Zhang2021_strain_skyrmion}.

The DC electric field holds promise for external controls of topological states of magnetic materials.
Historically, electric controls of magnetic systems have been discussed in the context of multiferroics~\cite{Tokura2014_review} and spintronics~\cite{Matsukura2015}.
For instance, noncollinear magnetic orders induce the electric polarization through a cross-correlation effect of the multiferroic material, that is, a magnetoelectric effect~\cite{Kimura2003_multiferroic,knb,Mostovoy2006,Dagotto2006,Cheong2007,Arima2007, Mostovoy2011_multiorbital,Tokura2014_review}.
For device applications, however, electric controls of the microscopic Hamiltonian is called for.
Figures~\ref{fig:e_vs_h}~(a) and (b) show schematic ground-state phase diagrams related to topological spin textures, which we will discuss later in Sec.~\ref{sec:Rashba}.
If we can increase the strength of the DMI $D(\bm E)$ by the external electric field $\bm E$, we will be able to generate topological spin textures by, for example, inducing the phase transition from the ferromagnetic phase to the the skyrmion-lattice phase.
To discuss such interesting phase transitions, we need a microscopic theory that gives us the strength of the DMI $D(\bm E)$ as a function of the DC electric field at the Hamiltonian level.

In addition, recent technological advances such as electric double-layer transistors~\cite{Bisri2017,ueno_iwasa_edlt_jpsj}, ferroelectric devices~\cite{Yazawa2021_APL,Mikolajick2021_JAP},
and interfacial engineering~\cite{Chen2013_interface,Matsuno2016_DM,kim2013_DM_interface,Huang2021_DMI_electric} make strong DC electric fields of $\sim1$--$10$~MV/cm available.
Scanning tunneling microscopes (STMs) can also yield DC electric fields of $\sim10$~MV/cm locally~\cite{chen2021_stm_book,Magtoto2000_stm}.
It is natural to expect that stronger electric fields change microscopic Hamiltonian more drastically.
Previously, the authors showed how the DC electric field quantitatively affects the microscopic superexchange coupling in magnetic Mott insulators~\cite{takasan_ex,furuya_superex}.
References~\cite{takasan_ex,furuya_superex} focus on magnetically isotropic cases.

On the other hand, concerning electric-field effects on magnetically anisotropic interactions, there are many previous theoretical and experimental studies~\cite{Mostovoy2006,Cheong2007,noh_dm_electric,jackeli_kitaev_2009,Khaliullin2005,knb}.
However, to the best of our knowledge, no microscopic theory  explicitly takes into account the geometrical configuration of ligand ions between magnetic ones to discuss electric-field effects on magnetic anisotropies.
Note that most of magnetic compounds have ligand ions between magnetic ions.
The ligand ion plays a crucial role in the exchange interaction between magnetic ions.
Hence, it is important to develop the microscopic theory that clarifies the role of ligand ions in DC electric-field controls of magnetic anisotropies.

This paper provides the general theoretical foundation to DC electric-field controls of magnetic anisotropy in Mott insulators by explicitly taking into account the ligand ion~\cite{Ohno2000,Romming2013, Matsukura2015,Chen2015,Matsuno2016_DM, Hsu2016,Jiang2018,Huang2018,Chen2019,Xu2020}.
Such a microscopic theory takes on a growing importance with ongoing advances in strong DC electric-field source~\cite{ueno_iwasa_edlt_jpsj,Bisri2017,Romming2013,Hsu2016}, including a single-cycle terahertz laser pulse~\cite{hirori_laser_2011,mukai_laser_2016,nicoletti_laser_2016,Fulop2019, sato_laser_dm,sato_floquet_book} (see Sec.~\ref{sec:estimation}). 
We consider two important cases: the Kitaev material and the magnetic Mott insulator with the Rashba SOC.
The former has the intra-atomic SOC that defines the most relevant orbitals to the quantum magnetism.
The latter has the inter-atomic SOC that will be generated on the interface of the target material to another one.
Our theory clarifies that the strong but feasible electric field can control Kitaev spin liquids and topological spin textures.

This paper is organized as follows.
We present a generic theoretical framework of our theory in Sec.~\ref{sec:framework}.
We then apply our theoretical framework to two specific cases.
The first case is the Kitaev spin liquid (Sec.~\ref{sec:Kitaev-Heisenberg}).
The second case is topological spin textures (Sec.~\ref{sec:Rashba}).
We summarize this paper and give discussions in Sec.~\ref{sec:discussions}.

\section{Framework}\label{sec:framework}

\begin{figure}[t!]
    \centering
    \includegraphics[bb = 0 0 1000 600, width=\linewidth]{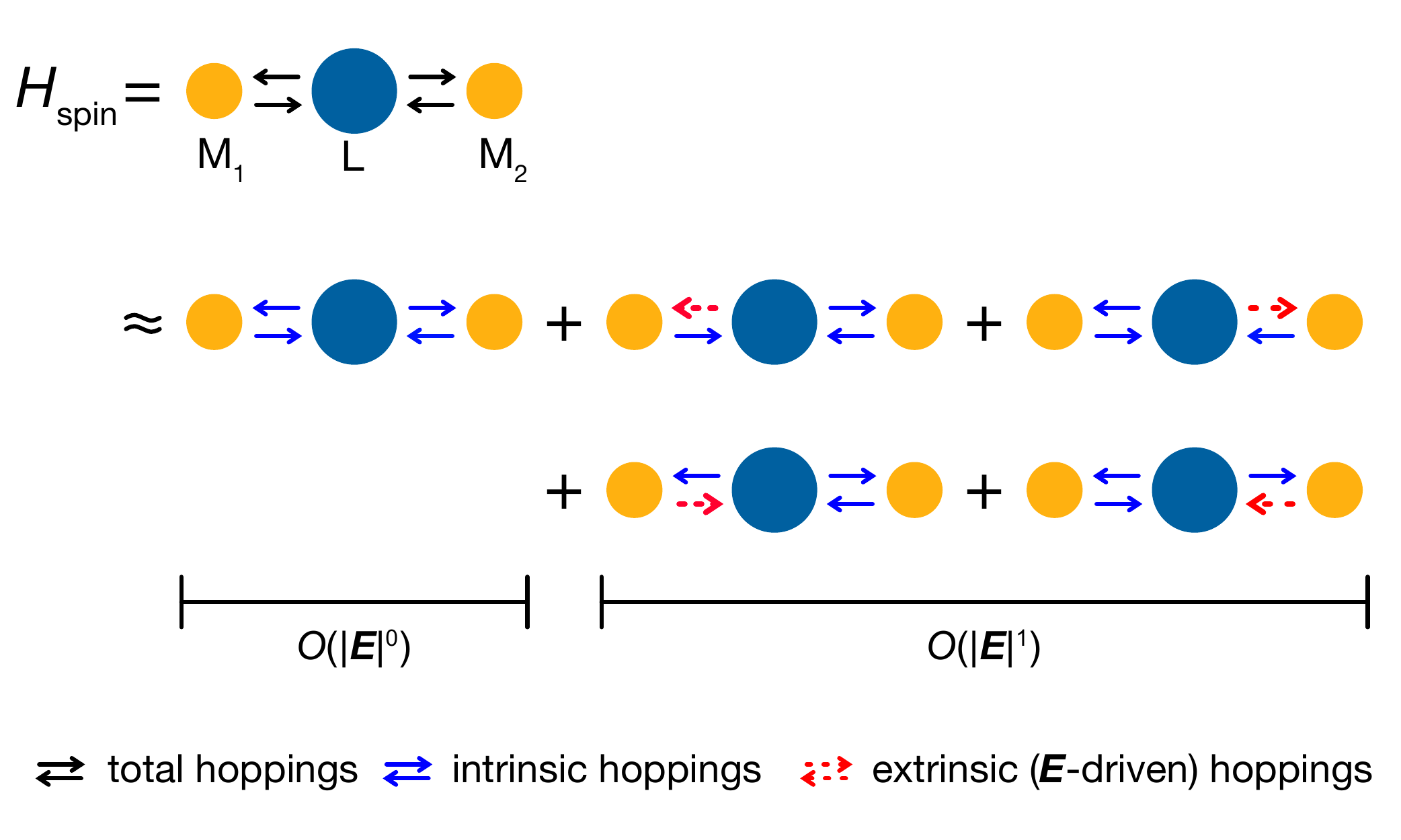}
    \caption{Diagrammatic representation of spin Hamiltonian \eqref{H_spin} with zeroth-order term omitted.
    The black solid, blue solid, and red dashed arrows denote the total hoppings $\mathcal{H}_t(\bm{E})$, the intrinsic hoppings $\mathcal{H}_t(\bm{0})$, and the $\bm{E}$-induced hoppings $\delta\mathcal{H}_t$, respectively.
    }
    \label{fig:diagrams}
\end{figure}

The spin Hamiltonian determining magnetic properties of Mott insulators crucially depends on virtual hopping processes of electrons under strong interactions.
These hopping processes are well described by Hubbard-like tight-binding models~\cite{moriya,eskes_superex,Khaliullin2005,jackeli_kitaev_2009,liu_khaliullin_2018,sugita_afm_kitaev_2020,furuya_superex}.
Considering the locality of virtual hopping processes in the Mott insulating phase, we derive a quantum spin system from the Hubbard-like model in two stages.
First, we consider a few-body electron system that contains two magnetic-ion sites (M$_1$ and M$_2$)  with half-occupied $d$ orbitals and a ligand site (L) with fully occupied $p$ orbitals.
This three-site model contains enough ingredients to yield low-energy spin-spin interactions between two neighboring magnetic-ion sites M$_1$ and M$_2$ unless further-neighbor electron hoppings become dominant.
Generally, the positional relation of the magnetic and ligand sites (such as an angle formed by M$_1$, M$_2$, and L) plays a critical role when determining the spin Hamiltonian under electric fields. In this paper, we focus on the isosceles triangle of Fig.~\ref{fig:gamma7_config}~(b).
Second, we build the many-body quantum spin model by combining many small blocks of the three-site model.
In this paper, we limit ourselves to $S=1/2$ quantum spin systems with superexchange interactions mediated by $p$ orbitals for simplicity.
Note that we can treat the $S=1/2$ spin and a pseudospin with the spin quantum number $1/2$  on equal footings in our framework.
In Sec.~\ref{sec:Kitaev-Heisenberg}, we consider a spin-orbit-entangled pseudospin, the $J_{\rm eff}=1/2$ doublet~\cite{kamimura_book,kim_jeff_1/2_2008,jackeli_kitaev_2009,onodera_group_1966,Takayama_review_spin-orbit_2021,Matsuura2013_JJ,Matsuura2014_compass}.

This paper deals with microscopic Hubbard-like models with the following generic Hamiltonian 
\begin{align}
    \mathcal{H}=\mathcal{H}_U(\bm{E})+\mathcal{H}_t(\bm{E}),
    \label{H_generic}
\end{align}
where $\mathcal{H}_U(\bm{E})$ and $\mathcal{H}_t(\bm{E})$ are intra-atomic and inter-atomic terms under electric fields $\bm{E}$, respectively.
Typically, the former contains on-site Coulomb repulsions and the latter does inter-atomic electron hoppings.
At zero electric fields, $\mathcal{H}_t(\bm{0})$ denotes the intrinsic electron hoppings and the difference,
\begin{align}
    \delta\mathcal{H}_t(\bm{E}):=\mathcal{H}_t(\bm{E})-\mathcal{H}_t(\bm{0}),
    \label{dHt_def}
\end{align}
contains extrinsic $\bm{E}$-induced hoppings.
We construct the Hubbard-like model \eqref{H_generic} so that it effectively describes low-energy physics of the target material under DC electric fields.
The field $\bm {E}$ may contain both the external electric field generated by the experimental instrument and an effective field generated on the interface of another material.
The detailed information of the material is encoded in operators and parameters of the model \eqref{H_generic}, as we will see in the next sections.
For example, the SOC and a crystal-electric field lift the orbital degeneracy and define orbitals relevant to low-energy quantum spin systems.
The model \eqref{H_generic} contains the information of relevant orbitals through the definition of the creation and annihilation operators in the second-quantized form.
The SOC also affects parameters included in the Hamiltonian \eqref{H_generic}.
The SOC can generate spin-flipping electron hoppings intrinsically or otherwise extrinsically in collaboration with the DC electric field $\bm{E}$.
Spin-flipping hoppings lead to magnetic anisotropies~\cite{moriya}.

We regard hoppings $\mathcal{H}_t(\bm{E})$ as a perturbation to $\mathcal{H}_U(\bm{E})$ because we are focused on the Mott-insulating phase induced by the on-site interaction $\mathcal H_U(\bm E)$.
Performing the fourth-order perturbation expansion, we obtain the effective spin Hamiltonian~\cite{furuya_superex},
\begin{align}
     \mathcal{H}_{\mathrm{spin}}&= P \mathcal{H}_UP+ P\mathcal{H}_t\biggl(\frac{1}{E_g-\mathcal{H}_U}Q\mathcal{H}_t\biggr)^3P,
     \label{H_spin}
\end{align}
where $P$ is the projection operator to the Mott-insulating ground states of $\mathcal{H}_U(\bm{E})$ with the eigenenergy $E_g$ and $Q=1-P$ is to its complementary space.
The zeroth-order term $P\mathcal{H}_U(\bm{E})P$ is mostly irrelevant but gives the uniform Zeeman energy, 
\begin{align}
    P\mathcal{H}_UP=-\sum_{a=x,y,z}\sum_jh^aS_j^a,
    \label{Zeeman_zeroth}
\end{align}
with $\bm{h}=\bm{g}\bm{B}$ for the magnetic field $\bm{B}=\mu_0\bm{H}$~\cite{furuya_superex}, where $\bm{g}$ is the electron's $g$ tensor.
Hereafter, we call $\bm{h}$ the magnetic field for simplicity.
$S_j^a$ is the $a$ component of the spin or pseudospin operator $\bm S_j = (S_j^x, S_j^y, S_j^z)$ for $a=x,y,z$. For details about the pseudospin, refer to Appendix~\ref{app:minimal}.
The second term of Eq.~\eqref{H_spin} gives the ligand-mediated superexchange interaction.

The perturbation $\mathcal H_t(\bm E)$ contains the $O(|\bm E|^0)$ term $\mathcal H_t(\bm 0)$ and
the $O(|\bm E|)$ term $\delta \mathcal H_t(\bm E)$.
It is reasonable to assume that the coupling constants in $\delta\mathcal H_t(\bm E)$ are much smaller than those in $\mathcal H_t(\bm 0)$ even under strong electric fields $\sim 1-10~\mathrm{MV/cm}$~\cite{furuya_superex}.
Hence, we approximate the effective spin Hamiltonian \eqref{H_spin} as follows.
\begin{widetext}
\begin{align}
    \mathcal H_{\rm spin}
    &\approx P \mathcal H_U P 
    + P \mathcal H_t(\bm 0) \biggl( \frac{1}{E_g-\mathcal H_U} Q\mathcal H_t(\bm 0) \biggr)^3 P 
    \notag \\
    &\qquad 
    + P\delta \mathcal H_t(\bm E) \biggl(\frac{1}{E_g-\mathcal H_U} Q \mathcal H_t(\bm 0) \biggr)^3 P + P \mathcal H_t(\bm 0 ) \frac{1}{E_g-\mathcal H_U} Q\delta \mathcal H_t(\bm E) \biggl(\frac{1}{E_g-\mathcal H_U} Q \mathcal H_t(\bm 0) \biggr)^2 P
    \notag \\
    &\qquad + P \mathcal H_t (\bm 0) \frac{1}{E_g-\mathcal H_U} Q \mathcal H_t(\bm 0)\frac{1}{E_g-\mathcal H_U} Q \delta \mathcal H_t(\bm E)\frac{1}{E_g-\mathcal H_U} Q \mathcal H_t(\bm 0) P \notag \\
    &\qquad + P \mathcal H_t(\bm 0) \biggl( \frac{1}{E_g-\mathcal H_U} Q \mathcal H_t(\bm 0) \biggr)^2 \frac{1}{E_g-\mathcal H_U} Q \delta \mathcal H_t(\bm E) P.
    \label{H_spin_approximate}
\end{align}
\end{widetext}
The first term of Eq.~\eqref{H_spin_approximate} represents the zeroth-order term \eqref{Zeeman_zeroth} of $\mathcal H_t(\bm E)$ and the other terms represent fourth-order terms of $\mathcal H_t(\bm E)$.
Each fourth-order term contains one $\delta \mathcal H_t(\bm E)$ at most since $\delta \mathcal H_t(\bm E)$ is proportional to $|\bm E|$.

Figure~\ref{fig:diagrams} shows a diagrammatic representation of these fourth-order processes in Eq.~\eqref{H_spin_approximate}.
The black, blue, and red arrows represent the total hoppings $\mathcal{H}_t(\bm{E})$, the intrinsic ones $\mathcal{H}_t(\bm{0})$, and the $\bm{E}$-induced ones, $\delta\mathcal{H}_t(\bm{E})$, respectively.
The intrinsic $O(|\bm{E}|^0)$ diagram typically contains the Heisenberg superexchange interaction~\cite{eskes_superex,furuya_superex}.
Hopping amplitudes of the $\bm{E}$-induced hoppings are $O(|\bm{E}|)$ because they arise from a distortion of electron orbitals' wave functions.
Note that our generic formalism deals with such distortions of wave functions in nonmagnetic ions as well as magnetic ones, as we will see later (e.g., see Sec.~\ref{sec:kitaev_in-plane}).
Importantly, the electric field can break a reflection symmetry.
We may rephrase the reflection-symmetry-breaking distortion as a locally generated $\bm{E}$-driven electric polarization.
The atomic-scale electric polarization triggers intrinsically forbidden hopping processes $\delta\mathcal{H}_t(\bm{E})$.
We obtain the $O(|\bm{E}|)$ correction to the spin Hamiltonian following the diagram.

The typical $O(|\bm E|)$ correction to the spin Hamiltonian is the DMI that breaks the inversion symmetry. The $\bm E$-induced DMI has been discussed in various contexts such as multiferroics~\cite{Kimura2003_multiferroic,knb,Mostovoy2006,Dagotto2006,Cheong2007,Arima2007, Mostovoy2011_multiorbital,Tokura2014_review}.
If the DMI is absent under zero electric fields, the coupling constant $\bm D(\bm E)$ of the $\bm E$-induced DMI $\bm D(\bm E) \cdot \bm S_i \times \bm S_j$ shows the following field dependence $D^a(\bm E)\approx g^{ab}E^b+\cdots$ with constant coefficients $g^{ab}$ for $a,b=x,y,z$.
Among many theoretical studies of the electric-field induction of magnetic interactions including the DMI, some studies adopt phenomenological symmetry arguments without deeply going into microscopic details~\cite{Mostovoy2006,Cheong2007,noh_dm_electric} and some others adopt second-order perturbation theory similar to ours~\cite{jackeli_kitaev_2009,Khaliullin2005,knb}.
As we briefly mentioned in Sec.~\ref{sec:introduction}, this paper adopts the fourth-order perturbation theory that explicitly takes into account the ligand ion and electron hoppings from and to the ligand site.
The explicit inclusion of the ligand site is the most characteristic point of this paper and makes it possible to clarify the microscopic origin of the coefficients $g^{ab}$ of the $\bm E$-induced DMI  (see e.g., Eq~\eqref{DF_parallel_kitaev}).
Besides, our argument is not limited to the DMI but also applicable to many other magnetic anisotropies.
We will show that the electric field can also yield the $\Gamma'$ interaction (see e.g., Eq.~\eqref{Gamma'_111}).

\section{Kitaev-Heisenberg model}\label{sec:Kitaev-Heisenberg}

\subsection{Introduction}

\begin{figure}[t!]
    \centering
    \includegraphics[bb = 0 0 450 300, width=\linewidth]{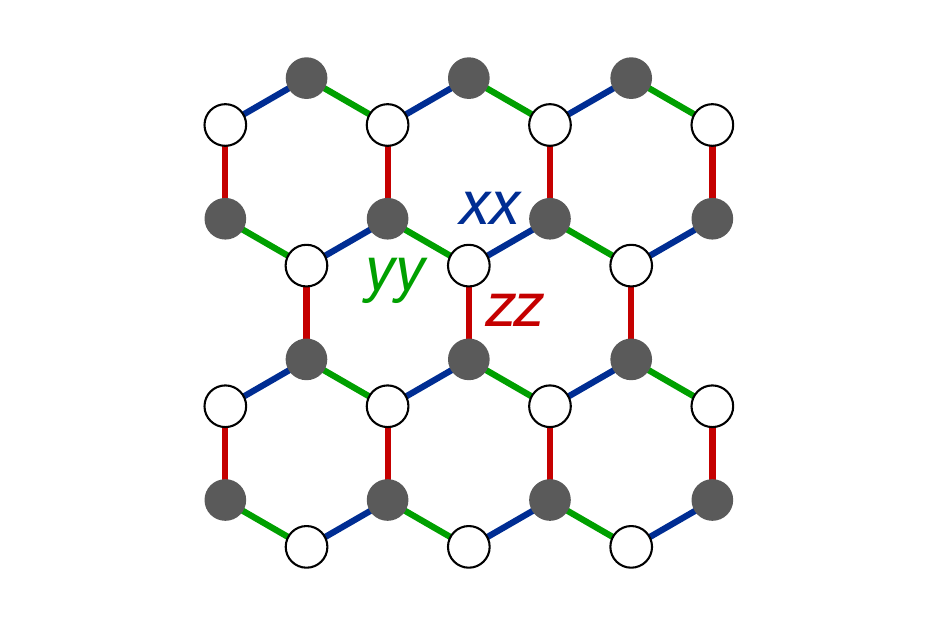}
    \caption{Kitaev interaction on honeycomb lattice.
    The blue ($xx$), green ($yy$), and red ($zz$) bonds denote the $\braket{i,j}_a$ bond in Eq.~\eqref{H_Kitaev} for $a=x,y,z$, respectively.
    }
    \label{fig:kitaev}
\end{figure}

\begin{figure}[t!]
    \centering
    \includegraphics[bb = 0 0 650 600, width=\linewidth]{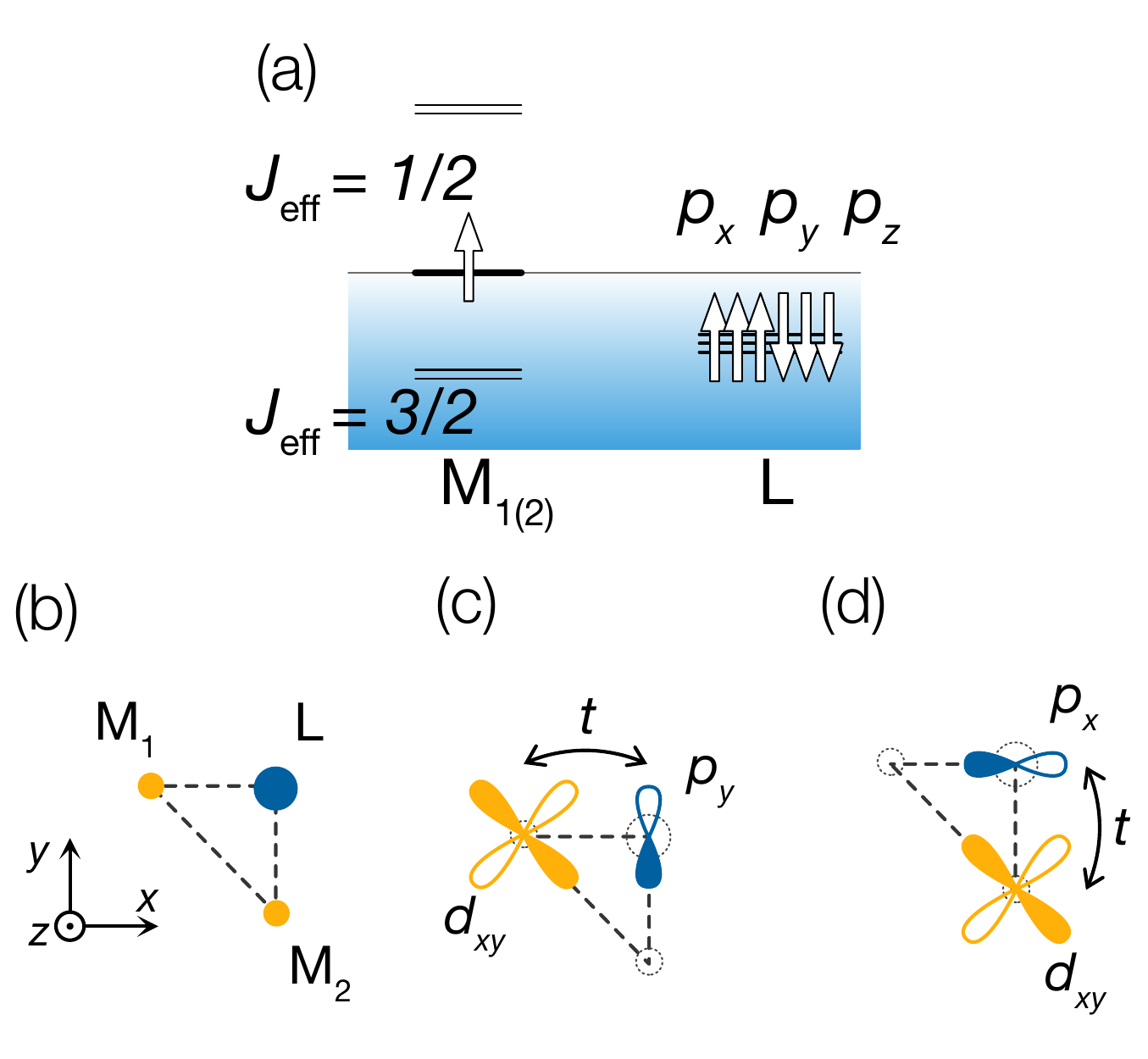}
    \caption{(a) $J_{\mathrm{eff}}=1/2$ model's electron configurations in $\Gamma_{7+}$ orbitals and $p_{x,y,z}$ orbitals at M$_j$ and L sites ($j=1,2$).
    The $J_{\rm eff}=1/2$ orbital is half occupied while the $J_{\rm eff}=3/2$ orbitals at the M$_j$ sites and the $p$ orbitals at the L site are fully occupied.
    (b) Geometrical configuration of three sites, M$_{1,2}$ and $L$. 
    $\triangle$M$_1$LM$_2$ forms a right-angle isosceles triangle.
    (c) Electron hopping between $d_{xy}$ orbital at M$_1$ and $p_y$ orbital at L.
    (d) Electron hopping between $d_{xy}$ orbital at M$_2$ and $p_x$ orbital at L.
    The hopping between $d_{xy}$ at M$_1$ (M$_2$) and $p_x$ ($p_y$) orbital at L is forbidden by symmetries.
    }
    \label{fig:gamma7_config}
\end{figure}

This section is devoted to the application of the generic theory to Kitaev materials~\cite{trebst_kitaev_review_2017,Motome_kitaev_review_2020,Hermanns_kitaev_review_2018}.
The Kitaev model is an $S=1/2$ quantum spin system on the honeycomb lattice with the Hamiltonian
\begin{align}
    \mathcal H_{\rm K} = \sum_{a=x,y,z} K^a \sum_{\braket{i,j}_a} S_i^a S_j^a,
    \label{H_Kitaev}
\end{align}
where $\braket{i,j}_a$ for $a=x,y,z$ denotes the $aa$ bond depicted in Fig.~\ref{fig:kitaev}.
$S_j^a$ denots the $a$ component of the $S=1/2$ spin operator $\bm S_j$ at the $j$th site.
The Hamiltonian \eqref{H_Kitaev} contains the Ising interaction $K^a S_i^a S_j^a$ whose direction depends on the bond $\braket{i,j}_a$.
The Kitaev model \eqref{H_Kitaev} has drawn much attention in the last decade for two reasons.
First, its ground state is exactly available and a spin liquid~\cite{kitaev}.
Second, Jackeli and Khaliulin showed possible realizations of the Kitaev model in spin-orbit-coupled Mott insulators~\cite{jackeli_kitaev_2009}.
Real materials have spin-spin interactions other than the Kitaev interactions (e.g., the Heisenberg interaction) however small they are.
There is a family of materials whose spin Hamiltonian contains the dominant Kitaev interaction and other subleading spin-spin interactions.
We call this family Kitaev materials in this paper.
For instance, the Kitaev-Heisenberg model 
\begin{align}
    \mathcal H_{\rm KH}
    &= \mathcal H_{\rm K} + J \sum_{\braket{i,j}} \bm S_i \cdot \bm S_j,
    \label{H_Kitaev_Heisenberg}
\end{align}
has often been discussed as a representative model of the Kitaev material, which may be more realistic than the Kitaev model \eqref{H_Kitaev}.
We can consider many other variants such as the Kitaev-Heisenberg-$\Gamma'$ model,
\begin{align}
    \mathcal H_{\rm KH\Gamma'} 
    &= \mathcal H_{\rm KH} + \Gamma'\sum_{a=x,y,z} \sum_{b,c\not=a}\sum_{\braket{i,j}_a} (S_i^bS_j^c+S_i^cS_j^b),
    \label{H_Kitaev_Heisenberg_Gamma'}
\end{align}
which we discuss later in this section.
This paper also deals with a variant of the Kitaev model that contains the electric-field-induced DMI.

In the rest of this section, we derive the Kitaev-Heisenberg model in the absence of the electric field and discuss how the electric field changes the model by adding extra spin-spin interactions.
We show that the electric field can yield the $\Gamma'$ interaction $\Gamma' (S_i^a S_j^b+S_i^bS_j^a)$ and the DMI.

\subsection{Without electric fields}

\subsubsection{Few-body system}

Based on the generic framework, we discuss Kitaev materials.
Here, we focus on a low-spin $d^5$ electron configuration under the octahedral crystal electric field and the strong SOC~\cite{jackeli_kitaev_2009}, where the $J_{\mathrm{eff}}=1/2$ doublet hosts a spin-orbit-entangled (pseudo)spin [Fig.~\ref{fig:gamma7_config}~(a)].
The $J_{\mathrm{eff}}^z=1/2$ state is a superposition of $t_{2g}$ states, $\ket{\uparrow,l^z=0}=\ket{d_{xy,\uparrow}}$ and $\ket{\downarrow,l^z=\pm 1}=(\ket{d_{zx,\downarrow}}\pm i \ket{d_{yz,\downarrow}})/\sqrt{2}$~\cite{kamimura_book,kim_jeff_1/2_2008,jackeli_kitaev_2009,onodera_group_1966,Takayama_review_spin-orbit_2021,Matsuura2013_JJ,Matsuura2014_compass}.
Let us denote the $J_{\rm eff}^z=\pm 1/2$ state on the $j$th magnetic ion as $\ket{\pm}_j$.
These pseudospin-1/2 states are written as
\begin{align}
    \ket{+}_j &= \frac{1}{\sqrt 3} (\ket{d_{j,xy,\uparrow}} + \ket{d_{j,yz,\downarrow}} +i \ket{d_{j,zx,\downarrow}}),
    \label{Jeff_+}
    \\
    \ket{-}_j &= \frac{1}{\sqrt 3} (\ket{d_{j,xy,\downarrow}} - \ket{d_{j,yz,\uparrow}} +i \ket{d_{j,zx,\uparrow}}).
    \label{Jeff_-}
\end{align}
We consider an electric configuration shown in Fig.~\ref{fig:gamma7_config}~(a).
One of $\ket{\pm}_j$ is occupied on the M$_j$ site and all the $p$ orbitals are occupied in the L site.

Let us consider a situation where the two magnetic ions M$_1$ and M$_2$ and the one nonmagnetic ion L~\footnote{Let us make a brief comment on the term ``ion'' in our context. Generally speaking, none of the M$_1$, M$_2$, or L sites needs to be ions. Practically, however, they are often ions that can release or receive electrons as assumed in the generic framework of Sec.~\ref{sec:framework}. Even though they could be other than ions, we call those sites are referred to as ``ions'' in this paper for notational simplicity.} form an isosceles right triangle $\triangle$M$_1$LM$_2$ on the $xy$ plane [Fig.~\ref{fig:gamma7_config}~(b)].
Figures~\ref{fig:gamma7_config}~(c) and (d) show examples of possible hoppings between $p$ and $d$ orbitals.
The $J_{\rm eff}=1/2$ doublet  is the superposition of $t_{2g}$ orbitals, as Eqs.~\eqref{Jeff_+} and \eqref{Jeff_-} show.
We first consider electron hoppings between $p$ and $t_{2g}$ orbitals and then rewrite the hoppings in terms of the $J_{\rm eff}=1/2$ doublet by projecting the $t_{2g}$ orbitals to the doublet.

It is straightforward to write the intrinsic hoppings $\mathcal{H}_t(\bm{0})$ in terms of $t_{2g}$-orbital operators.
As Figs.~\ref{fig:gamma7_config}~(c) and (d) show, the electron in the $p_y$ orbital can hop directly to the $d_{xy}$ orbital at M$_1$ but cannot 
hop directly to that at M$_2$ because the $d_{xy}$ orbital has the odd parity about a reflection $(x,y,z) \to (x,-y,z)$ but the $p_y$ orbital has the even parity.
Likewise, the electron in the $p_x$ orbital can hop directly to the $d_{xy}$ orbital at M$_2$ but cannot to that at M$_1$ because of the difference in the parity about another reflection $(x,y,z) \to (-x,y,z)$.
Such crystalline symmetries permit the following intrinsic hoppings at $\bm{E}=\bm{0}$:
\begin{align}
    \mathcal{H}_t(\bm{0})
    &= t \sum_{\sigma=\pm} (p_{y,\sigma}^\dag d_{1,xy,\sigma} + p_{z,\sigma}^\dag d_{1,zx,\sigma} 
    \notag \\
    &\qquad 
    + p_{x,\sigma}^\dag d_{2,xy,\sigma} + p_{z,\sigma}^\dag d_{2,zx,\sigma} + \mathrm{H.c.}).
    \label{Ht_KH_suppl}
\end{align}
The hopping amplitude $t$ represents a matrix element of a single-electron Hamiltonian,
\begin{align}
    H_1 = \frac{\bm p^2}{2m}+V(\bm x),
    \label{H_single}
\end{align}
where $V(x)$ is the potential that the electron feels, typically, a crystalline electric field.
For example, the hopping amplitude $t$ is given by
\begin{align}
    t=\braket{p_{y,\sigma}|H_1|d_{1,xy,\sigma}}.
\end{align}
Note that $t$ is independent of the index $\sigma$ because $H_1$ of Eq.~\eqref{H_single} has no $\sigma$ dependence.
In this section, we only consider the intra-atomic SOC and do not include any SOC in $H_1$.
When $H_1$ is independent of the electron spin $\sigma$, $t$ is independent of $\sigma$.
Other spatial symmetries of the triangle $\triangle$M$_1$LM$_2$ also removes the orbital dependence of $t$ in Eq.~\eqref{Ht_KH_suppl}.

\begin{figure}[t!]
    \centering
    \includegraphics[bb = 0 0 1200 700, width=\linewidth]{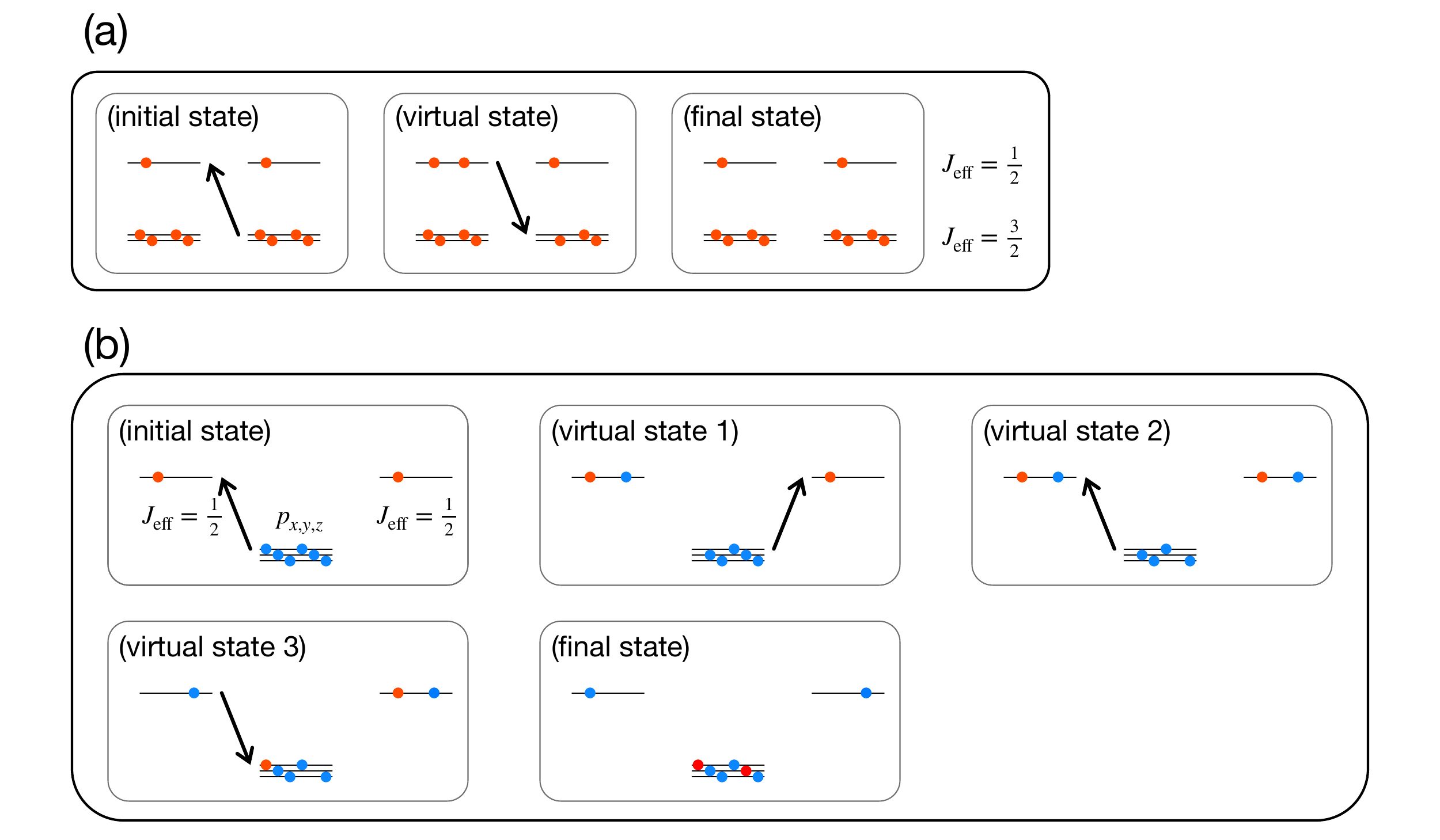}
    \caption{(a) Second-order process to yield Kitaev model discussed in Refs.~\cite{jackeli_kitaev_2009,bolens2018_kitaev,Matsuura2014_compass}.
    (b) Fourth-order process to yield Kitaev-Heisenberg model discussed in this paper.
    The arrows depict electron hoppings between proximate orbitals.
    }
    \label{fig:compass_vs_ours}
\end{figure}

The $J_{\rm eff}=1/2$ doublet [Eqs.~\eqref{Jeff_+} and \eqref{Jeff_-}] is the superposition of $d_{xy}$, $d_{yz}$, and $d_{zx}$ orbitals.
If we discard the $d$ orbitals other than the $J_{\rm eff}=1/2$ one, we can rewrite creation and annihilations operators of those $d$ orbitals 
are replaceable by those of $J_{\rm eff}=1/2$ ones $d_{j,\Gamma_{7+},\sigma}^\dag$ and $d_{j,\Gamma_{7+},\sigma}$ (see Appendix~\ref{app:zero-field_kitaev}):
\begin{align}
    \mathcal{H}_t(\bm{0})
    &=\frac{t}{\sqrt{3}}\sum_{\sigma=\pm} [(p_{y,\sigma}^\dag +ip_{z,-\sigma}^\dag)d_{1,\Gamma_{7+},\sigma}
    \notag \\
    &\qquad + (p_{x,\sigma}^\dag+\sigma p_{y,-\sigma}^\dag)d_{2,\Gamma_{7+},\sigma}+\mathrm{H.c.}].
    \label{Ht_Kitaev-Heisenberg}
\end{align}
Here, $\Gamma_{7+}$ labels the irreducible representation of the $J_{\mathrm{eff}}=1/2$ doublet~\cite{kamimura_book,onodera_group_1966}.

We are now ready to write down the full Hamiltonian
\begin{align}
    \mathcal H = \mathcal H_U + \mathcal H_t(\bm 0),
\end{align}
of the this three-site model at zero electric fields.
The intrinsic hoppings and the intra-atomic interactions $\mathcal H_U$ are given by Eq.~\eqref{Ht_Kitaev-Heisenberg} and
\begin{align}
	\mathcal H_U &= U_d \sum_{j=1,2} n_{j,\Gamma_{7+}, +} n_{j,\Gamma_{7+}, -} + U_p \sum_{\mu=x,y,z} n_{p_{\mu},+} n_{p_{\mu},-}
	\notag \\
	&\qquad 
	+ \sum_{j=1,2} V_j \sum_{\sigma=\pm} n_{j,\sigma} +V_p\sum_{\sigma=\pm} \sum_{\mu=x,y,z} n_{p_\mu,\sigma}
	\notag \\
	&\qquad
	-J_{\mathrm{H}} \sum_{\mu\not=\mu'} \bm s_{\mu}\cdot \bm s_{\mu'}-\sum_{j=1,2}\sum_{a=x,y,z}h^aS_j^a,
	\label{HU_KH_suppl}
\end{align}
respectively.
Here, $n_{j,\Gamma_{7+},\sigma}:=d_{j,\Gamma_{7+},\sigma}^\dag d_{j,\Gamma_{7+},\sigma}$ and
$n_{p_\mu,\sigma}:=p_{\mu,\sigma}^\dag p_{\mu,\sigma}$ are the electron number operators in the $J_{\rm eff}=1/2$ doublet at the M$_j$ site and $p_\mu$-orbital at the L site, respectively.
$U_d$ and $U_p$ denote the intra-band Coulomb repulsions for the $d$ and $p$ orbitals, $V_j$ and $V_p$ are the on-site potentials at M$_j$ and L, and $-J_{\mathrm{H}}<0$ is the ferromagnetic direct exchange between spins $\bm s_\mu$ in the $p_\mu$ orbitals.

Based on the generic framework of Fig.~\ref{fig:diagrams}, we obtain the following spin Hamiltonian from the intrinsic hoppings \eqref{Ht_Kitaev-Heisenberg} and the intra-atomic interactions:
\begin{align}
    \mathcal{H}_{\mathrm{spin}}(\bm{0})=J_{\mathrm{F}}\bm{S}_1\cdot\bm{S}_2+KS_1^zS_2^z-\sum_{a=x,y,z}\sum_{j=1,2}h^aS_j^a
    \label{H_Kitaev_Heisenberg}
\end{align}
with the exchange coupling, 
\begin{align}
    J_{\mathrm{F}}=-\frac{8}3t^4\biggl(\frac{1}{U_d-U_p+\Delta_{dp}}\biggr)^2\frac{1}{2(U_{d}-U_{p}+\Delta_{dp})-J_{\mathrm{H}}},
\end{align}
and the antiferromagnetic Kitaev coupling $K=-2J_{\mathrm{F}}>0$.
Here, $\bm S_j$ is the $J_{\rm eff}=1/2$ pseudospin  (see Appendix~\ref{app:minimal} for its definition) and $\Delta_{dp}=E_d-E_p$ is the difference of $J_{\rm eff}=1/2$-orbital eigenenergy ($E_d$) and the $p$-orbital one ($E_p$).
The coupling $J_{\rm F}$ is not necessarily ferromagnetic.
In this paper, we use parameters that yield the ferromagnetic coupling, $J_{\rm F}<0$, which is a typical case in accordance with the Goodenough-Kanamori rule~\cite{goodenough,kanamori1,kanamori2}.
Note also that the Kitaev coupling $K$ is independent of the SOC because we discarded the $J_{\mathrm{eff}}=3/2$ levels in the derivation of Eq.~\eqref{H_Kitaev_Heisenberg} [Fig.~\ref{fig:compass_vs_ours}~(b)].
Inclusion of the $J_{\mathrm{eff}}=3/2$ levels improves the quantitative aspect of the model, making $K$ depend on the SOC~\cite{Xu2018_npj_kitaev,Stavropoulos2019_prl_kitav}.
We discarded $J_{\rm eff}=3/2$ orbitals in this paper to keep the model as simple as possible.
By contrast, the explicit inclusion of the $p$ orbitals in our model is essential in order to discuss the electric-field effect microscopically~\cite{furuya_superex}.

\subsubsection{Many-body systems}

\begin{figure}
    \centering
    \includegraphics[bb = 0 0 1000 900, width=\linewidth]{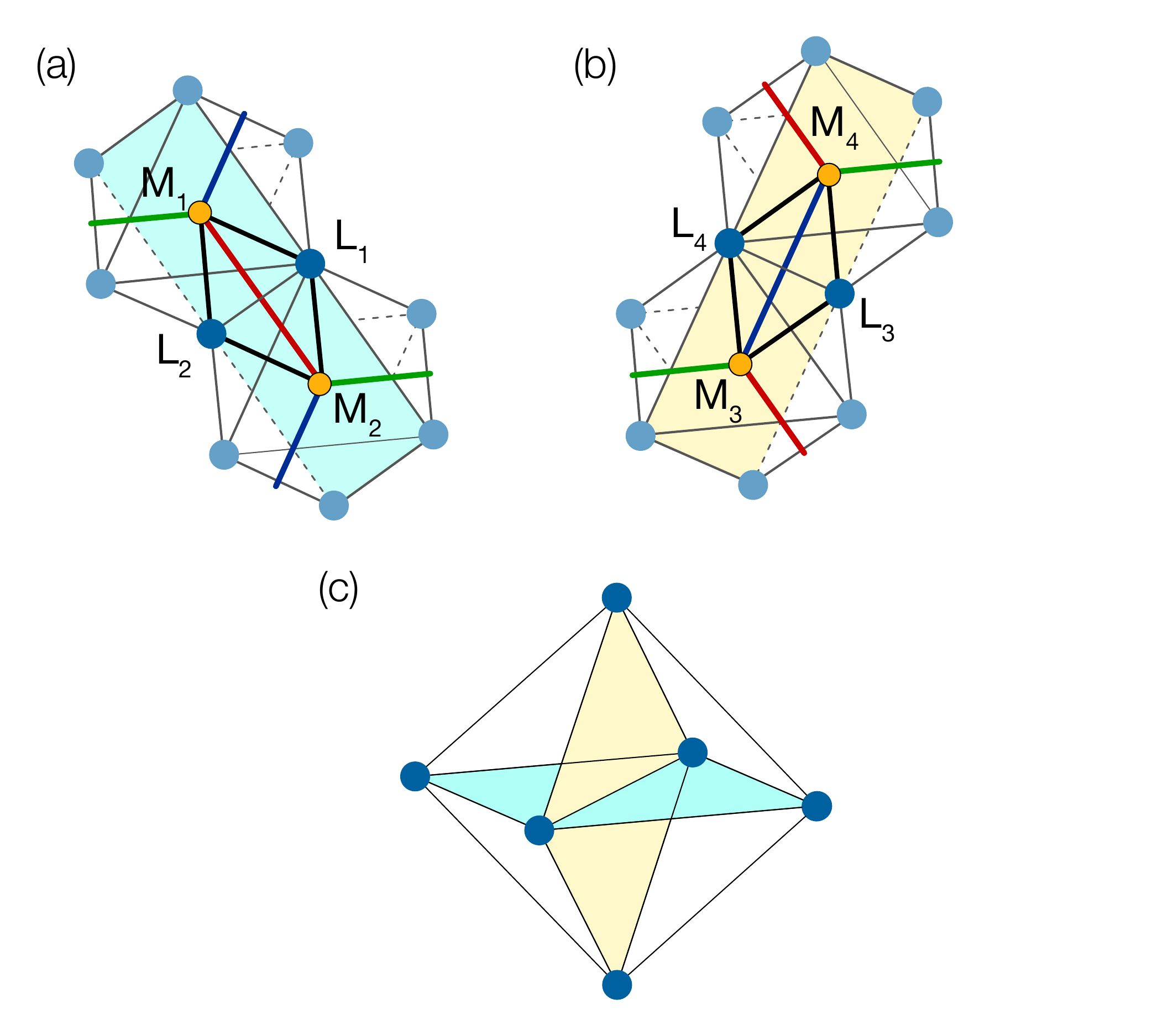}
    \caption{(a) $zz$ bond (red line) of Kitaev-Heisenberg model on $xy$ plane (light blue xxplane).
    (b) $xx$ bond (blue line) of Kitaev-Heisenberg model on $x'y'=yz$ plane (light yellow plane).
    (c) Relative positions of the two planes (a) and (b).
    }
    \label{fig:suppl_octahedra}
\end{figure}

We can build the Kitaev-Heisenberg model
\begin{align}
    \mathcal H_{\rm KH} &= \sum_{a=x,y,z} \sum_{\braket{i,j}_a} (KS_i^aS_j^a+J_{\mathrm{F}}\bm{S}_i\cdot\bm{S}_j)-\sum_j \bm{h}\cdot\bm{S}_j,
    \label{H_Kitaev-Heisenberg_suppl}
\end{align}
on the honeycomb lattice from the spin Hamiltonian \eqref{H_Kitaev_Heisenberg}.
Let us put two triangles $\triangle$M$_1$L$_1$M$_2$ and $\triangle$M$_1$L$_2$M$_2$ on the $xy$ plane as shown in Fig.~\ref{fig:suppl_octahedra}~(a).
Both the two triangles lead to the spin Hamiltonian \eqref{H_Kitaev_Heisenberg}.
The M$_1$--M$_2$ bond in this spin Hamiltonian corresponds to the $zz$ bond (the red bond) of Fig.~\ref{fig:kitaev}.
The other bonds $a=x,y$ are derived similarly.
Let us consider the $xx$ bond (the blue bond) of Fig.~\ref{fig:suppl_octahedra}~(b).
We consider a local $x'y'z'$ coordinate system and put triangles $\triangle$M$_3$L$_3$M$_4$ and $\triangle$M$_3$L$_4$M$_4$ on the $x'y'$ plane and derive the effective spin Hamiltonian,
\begin{align}
    \mathcal H_{\rm spin} &= P \biggl[KS_3^{z'}S_4^{z'} +J_{\mathrm{F}}\bm S_3 \cdot \bm S_4- \sum_{j=3,4}\bm h\cdot \bm S_j
    \biggr]P.
    \label{H_spin_x'y'}
\end{align}
To combine the octahedra of Figs.~\ref{fig:suppl_octahedra}~(a) and (b) so as to make them share an edge and form the green M$_2$--M$_3$ bond,
the $xyz$ and $x'y'z'$ coordinates should satisfy
\begin{align}
    (x',y',z')=(y,z,x),
    \label{120_rot_suppl}
\end{align}
turning Eq.~\eqref{H_spin_x'y'} into
\begin{align}
    \mathcal H_{\rm spin} &= P \biggl[KS_3^{x}S_4^{x} +J_{\mathrm{F}}\bm S_3 \cdot \bm S_4- \sum_{j=3,4}\bm h \cdot \bm S_j
    \biggr]P.
\end{align}
In fact, we can easily confirm  the relation \eqref{120_rot_suppl} by overlapping the $xy$ and $x'y'$ planes with a single octahedron [Fig.~\ref{fig:suppl_octahedra}~(c)].
Repeating the same procedure, we obtain the Kitaev-Heisenberg model \eqref{H_Kitaev-Heisenberg_suppl} on the honeycomb lattice of Fig.~\ref{fig:kitaev}
We may regard that the honeycomb lattice is put on the $(111)$ plane~\cite{jackeli_kitaev_2009}.

\subsection{With in-plane electric fields}\label{sec:kitaev_in-plane}

\begin{figure}[t!]
    \centering
    \includegraphics[bb = 0 0 600 360, width=\linewidth]{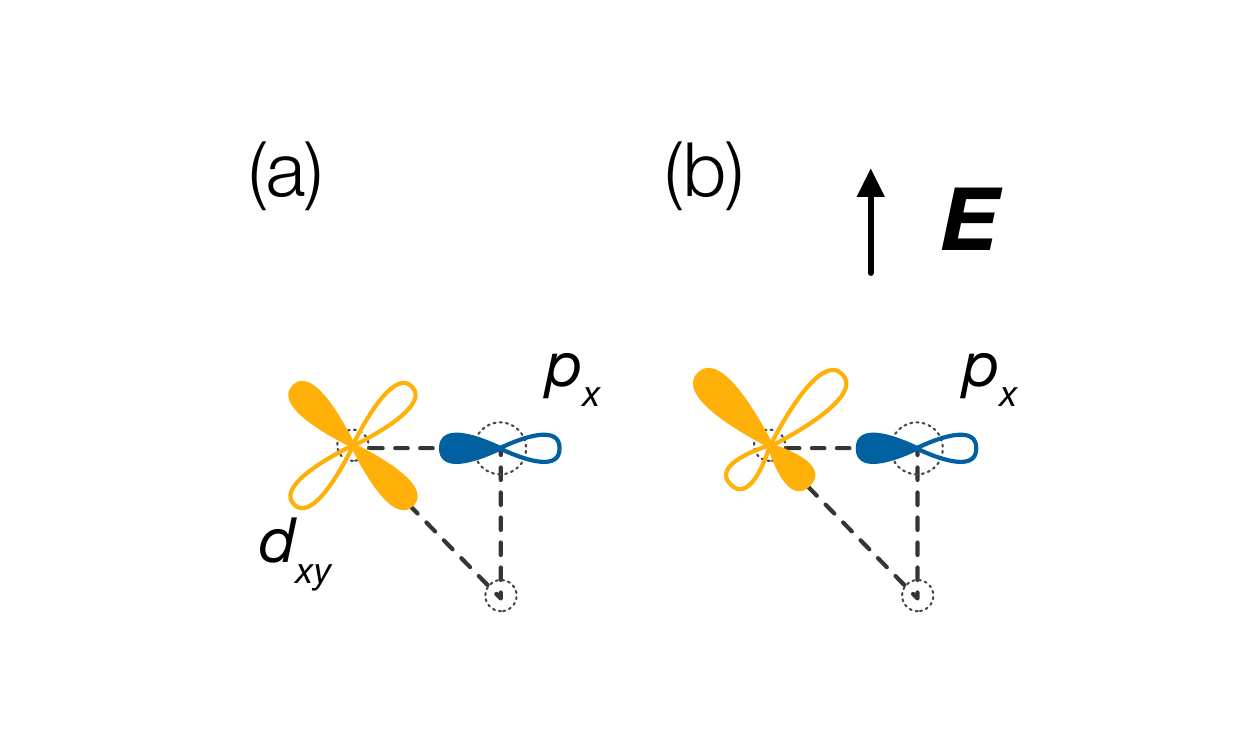}
    \caption{(a) Proximate $d_{xy}$ and $p_x$ orbitals with zero overlap integral. (b) $d_{xy}$ orbital distorted by electric field $\bm E$ along the vertical direction. This distortion makes the overlap of the two proximate orbitals nonzero, yielding the $\bm{E}$-induced hopping $\delta\mathcal H_t(\bm E)$.}
    \label{fig:dist_suppl}
\end{figure}

\begin{figure}[t!]
    \centering
    \includegraphics[bb = 0 0 1000 900, width=\linewidth]{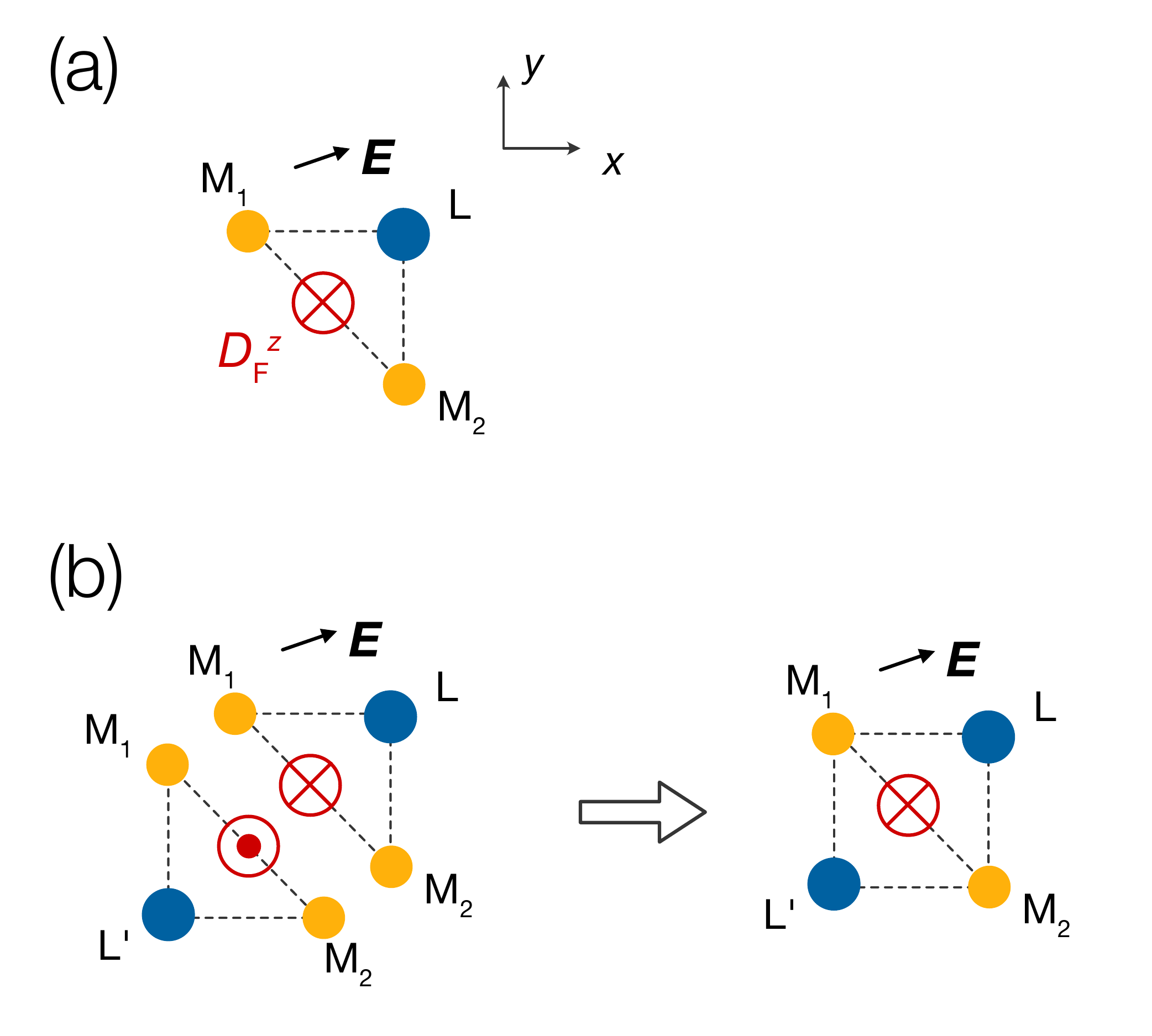}
    \caption{
    (a) Out-of-plane DMI in $J_{\mathrm{eff}}=1/2$ model induced by in-plane $\bm{E}$.
    (b) Combining two triangles $\triangle$M$_1$LM$_2$ and $\triangle$M$_1$L$'$M$_2$ into a rectangle M$_1$LM$_2$L$'$, we obtain the DMI $D_{\rm F}^z\bm e_z$ and $D'^z_{\rm F} \bm e_z$, respectively.
    The resultant DMI on the M$_1$-M$_2$ bond of the rectangle is $(D_{\rm F}^z+D'^z_{\rm F})\bm e_z$.
    If L and L$'$ are equivalent, $D'^z_{\rm F}=-D_{\rm F}^z$ follows from the reflection M$_1\leftrightarrow$ M$_2$ about the $x=y$ plane.
    }
    \label{fig:kitaev_DMI_in-plane-field}
\end{figure}

We go back to the three-site model on the $xy$ plane (i.e., $z=0$) and now give our attention to $\bm{E}$-induced interactions that are to be added to the spin Hamiltonian \eqref{H_Kitaev_Heisenberg}.
Let us apply the in-plane electric field $\bm{E}=E^x\bm{e}_x+E^y\bm{e}_y$ to our three-site model on $\triangle$M$_1$LM$_2$.
The in-plane electric field breaks the reflection symmetry $(x,y,z)\to(y,x,z)$ about the $x=y$ plane.
The other reflection symmetry $(x,y,z)\to(x,y,-z)$ about the $z=0$ plane remains intact.
The in-plane electric field permits hoppings that were intrinsically forbidden.
Lowering the spatial symmetry of the system, the in-plane DC electric field induces the following hoppings between $p$ orbitals and $t_{2g}$ orbitals,
\begin{align}
    \delta \mathcal{H}_t(\bm{E})
    &= -I \sum_{\sigma=\pm} [E^y (p_{z,\sigma}^\dag d_{1,yz,\sigma}+ p_{x,\sigma}^\dag d_{1,xy,\sigma}) 
    \notag \\
    &\qquad 
    + E^x (p_{z,\sigma}^\dag d_{2,yz,\sigma} +
    p_{y,\sigma}^\dag d_{2,xy,\sigma}) + \mathrm{H.c.}
    ],
\end{align}
where $-IE^y$ is the matrix element of the  $-E^yP^y$ term in the single-electron Hamiltonian:
\begin{align}
    -IE^y &=\braket{p_{x,\sigma}| (-E^yP^y)|d_{1,xy,\sigma}}
    \notag \\
    &= -e E^y \int d\bm{r} \braket{p_{x,\sigma}|\bm r}\, y\, \braket{\bm r|d_{j,xy,\sigma}}.
    \label{IEy_mixture_suppl}
\end{align}
The matrix element \eqref{IEy_mixture_suppl} represents a reflection-symmetry-breaking distortion of the wave functions of $d_{1,xy,\sigma}$ and $p_{x,\sigma}$ orbitals caused by the electric field.
The perturbation $V=-\bm{E}\cdot\bm{P}$ gives rise to 
the following perturbation correction to $\ket{d_{1,xy,\sigma}}$:
\begin{align}
    \ket{d_{1,xy,\sigma}}  \to  \ket{d_{1,xy,\sigma}} - \frac{\braket{p_{x,\sigma}|V|d_{1,xy,\sigma}}}{E_p-E_d}\ket{p_{x,\sigma}}  +\cdots
\end{align}
The similar expansion holds for $\ket{p_{x,\sigma}}$.
The matrix element $\braket{p_{x,\sigma}|V|d_{1,xy,\sigma}}=-IE^y$ gives a measure of how much the  proximate $d_{xy}$ and $p_x$ orbitals mix with each other.
Their mixture results in the distortion of orbital probability clouds along the direction of $\bm{E}$ [from Fig.~\ref{fig:dist_suppl}~(a) to Fig.~\ref{fig:dist_suppl}~(b)].
The distortion of the orbitals affects the spin Hamiltonian as the addition of the $\bm{E}$-induced hoppings $\delta \mathcal H_t(\bm E)$.
The additional hoppings keep the reflection symmetry about the $z=0$ plane in accordance with the direction of the in-plane electric field.
This reflection symmetry permits the DMI with the nonzero $z$ component and with zero $x$ and $y$ components [Fig.~\ref{fig:kitaev_DMI_in-plane-field}~(a)].

To derive the $\bm{E}$-induced DMI, we first rewrite the hoppings $\delta\mathcal H_t(\bm E)$ similarly to the intrinsic hoppings $\mathcal H_t(\bm{0})$:
\begin{align}
    \delta \mathcal H_t(\bm{E})
    &=  -\frac{I}{\sqrt 3} \sum_{\sigma=\pm} [E^y (\sigma p_{z,-\sigma}^\dag +
    p_{x,\sigma}^\dag) d_{1,\Gamma_{7+},\sigma}
    \notag \\
    &\qquad + E^x (\sigma p_{z,-\sigma}^\dag + p_{y,\sigma}^\dag) d_{2,\Gamma_{7+},\sigma}
    \notag \\
    &\qquad + \mathrm{H.c.}),
\end{align}
where spin-dependent hoppings (e.g., $p_{z,-\sigma}^\dag d_{1,\Gamma_{7+},\sigma}$) emerged because the $J_{\rm eff}=1/2$ doublet \eqref{Jeff_+} and \eqref{Jeff_-} contain spin-up and spin-down states of $t_{2g}$ orbitals.
For example, $\ket{+}_j$ contains $\ket{d_{j,xy,\uparrow}}$ and $\ket{d_{j,zx,\downarrow}}$.
Straightforward calculations of Eq.~\eqref{H_spin_approximate} show that the in-plane electric field adds 
the following DMI
\begin{align}
    \bm{D}_{\mathrm{F}}\cdot\bm{S}_1\times\bm{S}_2=D_{\mathrm{F}}^z(\bm{S}_1\times\bm{S}_2)^z,
    \label{DM_in_plane}
\end{align}
with 
\begin{align}
    \frac{D_{\mathrm{F}}^z}{J_{\mathrm{F}}}=-\frac{4I}{t}\frac{(E^x+E^y)(U_d-U_p+\Delta_{dp})+(E^x-E^y)J_{\mathrm{H}}}{2(U_d-U_p)+J_{\mathrm{H}}},
\end{align}
to the Kitaev-Heisenberg Hamiltonian \eqref{H_Kitaev_Heisenberg}, which is in accordance with the reflection symmetry $(x,y,z) \to (x,y,-z)$ that the in-plane electric field respects.
As we did in the previous subsection, we combine $\triangle$M$_1$LM$_2$ and $\triangle$M$_1$L$'$M$_2$ into the rectangle M$_1$LM$_2$L$'$ [Fig.~\ref{fig:kitaev_DMI_in-plane-field}~(b)]. 
Let us suppose that the DMI generated by the in-plane $\bm E$ on these triangles are $D_{\rm F}^z(\bm S_1\times \bm S_2)^z$ and $D'^z_{\rm F}(\bm S_1\times \bm S_2)^z$ as shown in Fig.~\ref{fig:kitaev_DMI_in-plane-field}~(b).
The DMI on the M$_1$-M$_2$ bond on the rectangle M$_1$LM$_2$L$'$ is their superposition, $(D_{\rm F}^z+D'^z_{\rm F})(\bm S_1\times \bm S_2)^z$.
Note that if L and L$'$ are completely equivalent, $D'^z_{\rm F} = -D_{\rm F}^z$ follows from the equivalence.
The DMI is then absent in the rectangle and eventually in the honeycomb lattice (Fig.~\ref{fig:kitaev}).
The DM vector $\bm D_{\rm F}$ of the $\bm E$-induced DMI~\eqref{DM_in_plane} is perpendicular to  $\triangle$M$_1$LM$_2$. Note that this direction of the DM vector perfectly accords with Moriya's symmetry argument~\cite{moriya} about the direction of the DM vector.
The in-plane electric field keeps the inversion symmetry about the $xy$ plane: $(x,y,z)\to (x,y,-z)$.
This inversion forbids the $x$ and $y$ components of the DMI.
Indeed, the inversion transforms the DM vector as $(D_{\rm F}^x(\bm E), D_{\rm F}^y(\bm E), D_{\rm F}^z(\bm E)) \to (-D_{\rm F}(\bm E), \, -D_{\rm F}(\bm E), \, D_{\rm F}(\bm E))$, where $\bm E$ is the in-plane electric field.
The inversion symmetry about the $xy$ plane thus leads to $D_{\rm F}^x=D_{\rm F}^y=0$, in accordance with the microscopically derived result \eqref{DM_in_plane}.

\subsection{With out-of-plane electric fields}

\begin{figure}[t!]
    \centering
    \includegraphics[bb = 0 0 1000 900, width=\linewidth]{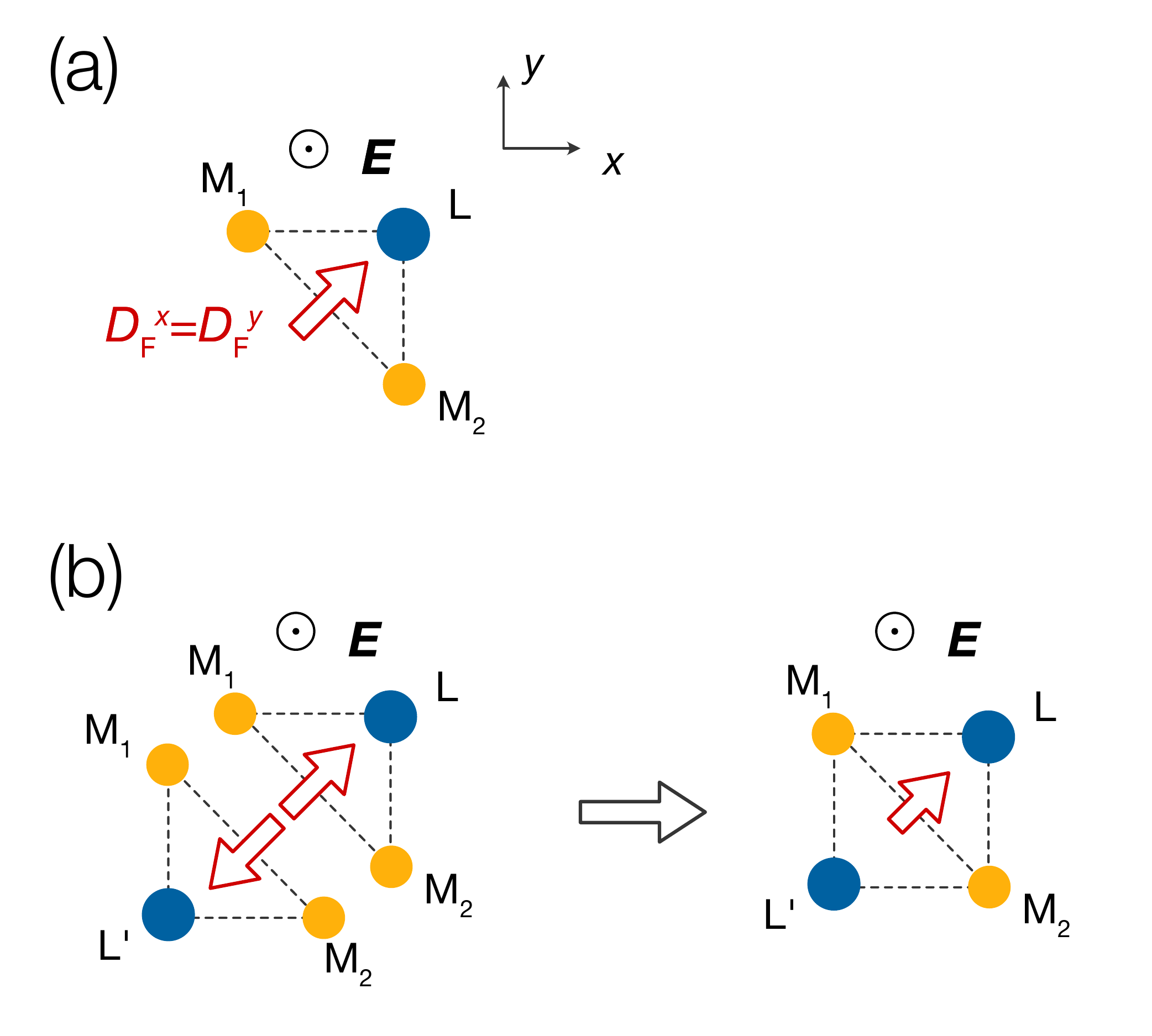}
    \caption{
    (a) In-plane DMI in $J_{\mathrm{eff}}=1/2$ model induced by out-of-plane $\bm{E}$.
    The DMI has the DM vector $D_{\rm F}^x\bm e_x + D_{\rm F}^y \bm e_y$ with $D_{\rm F}^x=D_{\rm F}^y=: D_{\rm F}^\parallel$.
    (b) Combining two triangles $\triangle$M$_1$LM$_2$ and $\triangle$M$_1$L$'$M$_2$ into a rectangle M$_1$LM$_2$L$'$, we obtain the DMI $D_{\rm F}^\parallel (\bm e_x+\bm e_y)$ and $D'^\parallel_{\rm F} (\bm e_x+\bm e_y)$, respectively.
    The resultant DMI on the M$_1$-M$_2$ bond of the rectangle is $(D_{\rm F}^\parallel+D'^\parallel_{\rm F}) (\bm e_x+\bm e_y)$.
    If L and L$'$ are equivalent, $D'^\parallel_{\rm F}=-D_{\rm F}^\parallel$ follows from the reflection M$_1\leftrightarrow$ M$_2$ about the $x=y$ plane.
    }
    \label{fig:kitaev_DMI_out-of-plane-field}
\end{figure}

Let us apply an out-of-plane electric field $\bm E=E^z\bm e_z$ to the triangle $\triangle$M$_1$LM$_2$ on the $z=0$ plane.
The out-of-plane electric field violates the reflection symmetry about the $z=0$ plane but keeps
a discrete rotational symmetry, a $\pi$-rotational one about a perpendicular bisector $\ell$ of the triangle $\triangle\,\mathrm{M_1LM_2}$.
Note that the red-fringe arrow of Fig.~\ref{fig:kitaev_DMI_out-of-plane-field}~(a) is on the line $\ell$.
The line $\ell$ is defined as $\ell=\{(x,y,z)| \, y=x, \, z=0\}$.
The $\pi$ rotation around $\ell$ changes $(x,y,z)\to (y,x,-z)$.
Let us denote the $\bm E$-induced DM vector due to $\bm E=E^z\bm e_z$ as $\bm D_{\rm F}(E^z)$.
The $\pi$ rotation $(x,y,z)\to (y,x,-z)$ affects the DMI as follows.
\begin{align}
    &(D^x(E^z), \, D^y(E^z), \, D^z(E^z) ) 
    \notag \\
    &\qquad \to (-D^x(-E^z), \, -D^y(-E^z), \, D^z(-E^z)).
    \label{DMI_symmetry_out-of-plane-E}
\end{align}
As we did in the previous subsection, we evaluate the DMI perturbatively.
If we consider the linear effect of $\bm E$ only, the DM vector is approximately an odd function of the out-of-plane field $E^z$: $\bm D_{\rm F}(-E^z)=- \bm D_{\rm F}(E^z)$.
Therefore, Eq.~\eqref{DMI_symmetry_out-of-plane-E} is approximated as
\begin{align}
    &(D^x(E^z), \, D^y(E^z), \, D^z(E^z) ) 
    \notag \\
    &\qquad \to (D^x(E^z), \, D^y(E^z), \, -D^z(E^z)).
    \label{DMI_symmetry_out-of-plane-E_1st_approx}
\end{align}
Equation~\eqref{DMI_symmetry_out-of-plane-E_1st_approx} implies that the $\bm E$-induced DMI shows $D_{\rm F}^x=D_{\rm F}^y$ and $D_{\rm F}^z=0$ with the linear order of $\bm E$.

Let us derive the DMI microscopically without relying on the symmetry argument.
The out-of-plane electric field induces the following hopping:
\begin{align}
    \delta \mathcal H_t(\bm E)
    &= - IE^z \sum_{\sigma=\pm}\sum_{j=1,2} [p_{y,\sigma}^\dag   d_{j,yz,\sigma} + p_{x,\sigma}^\dag d_{j,zx,\sigma}
    +\mathrm{H.c.}]
    \notag \\
    &= - \frac{IE^z}{\sqrt 3} \sum_{\sigma,j} [ (\sigma p_{y,-\sigma}^\dag + i p_{x,-\sigma}^\dag ) d_{j,\Gamma_{7+},\sigma}+\mathrm{H.c.}
    ].
    \label{dHt_out-of-plane}
\end{align}

Repeating the same procedure as the one in the previous subsection,
we straightforwardly obtain
\begin{align}
    \mathcal{H}_{\mathrm{spin}}&=J_{\mathrm{F}}\bm{S}_1\cdot\bm{S}_2 +KS_1^zS_2^z-\sum_{a=x,y,z}\sum_{j=1,2}h^aS_j^a
    \notag \\
    &+\bm{D}_{\mathrm{F}}\cdot(\bm{S}_1\times\bm{S}_2)+ \Gamma'[(S_1^x+S_1^y) S_2^z+S_1^z (S_2^x+S_2^y)],
    \label{H_spin_kitaev_out-of-plane}
\end{align}
with
\begin{align}
    \bm{D}_{\mathrm{F}}
    &=D_{\mathrm{F}}^\parallel(\bm{e}_x+\bm{e}_y), \\
    D_{\mathrm{F}}^\parallel
    &=\frac{16t^3}{3}E^zI\biggl(\frac{1}{U_d-U_p+\Delta_{dp}}\biggr)^2
    \notag \\
    &\qquad \times \frac{J_{\mathrm{H}}}{4(U_d-U_p+\Delta_{dp})^2-{J_{\mathrm{H}}}^2}, 
    \label{DF_parallel_kitaev}
    \\
    \Gamma'
    &=\frac{32t^3}{9}E^zI\biggl(\frac{1}{U_d-U_p+\Delta_{dp}}\biggr)^2
    \notag \\
    &\qquad \times \frac{U_d-U_p+\Delta_{dp}}{4(U_d-U_p+\Delta_{dp})^2-{J_{\mathrm{H}}}^2}.
\end{align}
The out-of-plane electric field gives the DMI and the inversion-symmetric off-diagonal magnetic anisotropy,
$\Gamma'(S_1^yS_2^z+S_1^zS_2^y)$, known as the $\Gamma'$ interaction~\cite{takikawa_kitaev_off-diagonal_2019,takikawa_kitaev_off-diagonal_2020} in the context of Kitaev materials.
Both the DMI and the $\Gamma'$ interaction accord with the reflection symmetry about the $x=y$ plane.

When we combine two triangles $\triangle$M$_1$LM$_2$ and $\triangle$M$_1$L$'$M$_2$ into the rectangle M$_1$LL$'$M$_2$ in analogy with the in-plane $\bm E$ case, we face the same cancellation of the DMI again.
If L and L$'$ are equivalent, $D_{\rm F1}^\parallel$ of $\triangle$M$_1$LM$_2$ and $D_{\rm F2}^\parallel$ of $\triangle$M$_1$L$'$M$_2$ have the same magnitude and the opposite sign: $D_{\rm F1}^\parallel = -D_{\rm F2}^\parallel$ [Fig.~\ref{fig:kitaev_DMI_out-of-plane-field}~(b)].
Hence, the M$_1$--M$_2$ bond of the rectangle M$_1$LL$'$M$_2$ has then no DMI after all.
The relation $D_{\rm F1}^\parallel=-D_{\rm F2}^\parallel$ directly comes from the fact that the DMI is antisymmetric under the reflection about the $x=y$ plane.
By contrast, the $\Gamma'$ interaction is symmetric under that reflection.
Therefore, the $\Gamma'$ interaction is present on the M$_1$--M$_2$ bond of the combined rectangle.

To conclude this subsection, we note that the spin Hamiltonian \eqref{H_spin_kitaev_out-of-plane} indeed accords with the $\pi$-rotational symmetry about the line $\ell$.
As Eq.~\eqref{DMI_symmetry_out-of-plane-E_1st_approx} shows, the $\bm E$-induced DMI lies in the $xy$ plane (i.e., $D_{\rm F}^z=0$) within our theoretical scheme. 
Note that the out-of-plane component $D_{\rm F}(E^z)$ is not forbidden but a higher-order effect of the electric field, practically negligible.
The same argument applies to the $\Gamma'$ interaction.
Let us consider a generic $\Gamma'$ interaction:
\begin{align}
    &\Gamma'_x(E^z) (S_1^yS_2^z+S_1^zS_2^y) + \Gamma'_y(E^z) (S_1^zS_2^x+S_1^xS_2^z)
    \notag \\
    &\qquad + \Gamma'_z(E^z) (S_1^xS_2^y+S_1^yS_2^x).
\end{align}
The $\pi$ rotation changes 
\begin{align}
    &(\Gamma'_x(E^z), \, \Gamma'_y(E^z),\, \Gamma'_z(E^z))
    \notag \\
    &\qquad \to (-\Gamma'_y(-E^z), \, -\Gamma'_x(-E^z),\, \Gamma'_z(-E^z)).
\end{align}
If we keep the $O(\bm E)$ terms only, we can approximate this transformation as
\begin{align}
    &(\Gamma'_x(E^z), \, \Gamma'_y(E^z),\, \Gamma'_z(E^z))
    \notag \\
    &\qquad \to (\Gamma'_y(E^z), \, \Gamma'_x(E^z),\, -\Gamma'_z(E^z)).
\end{align}
In other words, the $\bm E$-induced $\Gamma'$ interaction satisfies $\Gamma'_x(E^z)=\Gamma'_y(E^z)$ and $\Gamma'_z(E^z)=0$, as the microscopically derived spin Hamiltonian \eqref{H_spin_kitaev_out-of-plane} shows.

\subsection{Kitaev-Heisenberg-$\Gamma'$ model under [111] electric field}

Recall that the honeycomb lattice of Fig.~\ref{fig:kitaev} is put on the $(111)$ plane.
Now we apply the electric field $\bm E_{[111]}=E_{[111]}(\bm e_x+\bm e_y+\bm e_z)/\sqrt{3}$ so that the electric field is perpendicular to the honeycomb lattice.
This electric field $\bm E_{[111]}$ has both the in-plane and out-of-plane components.
Within our framework (Fig.~\ref{fig:diagrams}) that includes the $\bm E$-induced hopping up to the linear order of $|\bm E|$,
the $\bm E_{[111]}$-induced magnetic anisotropies are a simple superposition of those induced by the in-plane field and the out-plane-plane ones.
Hence, we obtain the following spin Hamiltonian,
\begin{align}
    \mathcal H_{\mathrm{KH}\Gamma'}
    &= \sum_{a=x,y,z}\sum_{\braket{i,j}_a} (K S_i^a S_j^a + J_{\mathrm{F}}\bm S_i \cdot \bm S_j) 
    - \sum_j \bm h \cdot \bm S_j
    \notag \\
    &\qquad +\Gamma'(\bm E_{[111]}) \sum_{a=x,y,z} \sum_{b,c\not=a} \sum_{\braket{i,j}_a} (S_i^bS_j^c+S_i^cS_j^b),
    \label{H_KHgamma'}
\end{align}
where $\Gamma'(\bm E_{[111]})$ is proportional to $E_{[111]}$:
\begin{align}
    \Gamma' &=\frac{32t^3}{9}\frac{2E_{[111]}}{\sqrt 3}I\biggl(\frac{1}{U_d-U_p+\Delta_{dp}}\biggr)^2
    \notag\\
    &\qquad \cdot \frac{U_d-U_p+\Delta_{dp}}{4(U_d-U_p+\Delta_{dp})^2-{J_{\mathrm{H}}}^2}.
    \label{Gamma'_111}
\end{align}
We assumed that the ligand sites are all equivalent.
As we saw in the previous subsections, the DMI induced by $\bm E_{[111]}$ is completely canceled when L and L$'$ are equivalent in the building blocks [Figs.~\ref{fig:suppl_octahedra}~(a) and (b)] of the honeycomb lattice (Fig.~\ref{fig:kitaev}).
When the L and L' sites are nonequivalent, the effective spin Hamiltonian \eqref{H_KHgamma'} further acquires the DMI.
The spin Hamiltonian \eqref{H_KHgamma'} thus contains the Kitaev interaction, the Heisenberg interaction, the Zeeman energy, and the $\Gamma'$ interaction.
We call it a Kitaev-Heisenberg-$\Gamma'$ model in this paper.

\subsection{Gapped quantum spin liquid}\label{sec:gapped_QSL}

Much effort is being made to investigate effects of non-Kitaev interactions on quantum spin liquid states of the pure Kitaev model in connection with Kitaev materials such as $\alpha$-RuCl$_3$~\cite{Banerjee2016_rucl3,Yadav2016_rucl3,glamazda2017_rucl3,wolter2017_rucl3,Kasahara2018_rucl3,Wulferding2020_rucl3,Yamashita2020_rucl3,Suzuki2021_rucl3}.

Under uniform magnetic fields $\bm h=\sum_{a=x,y,z}h^a\bm e_a$, the Kitaev model can have a gapped quantum spin liquid phase with the Chern number $C=1$ when $h^xh^yh^z\not=0$~\cite{Kitaev_2006}.
The Majorana fermion in that phase has a mass gap $m$, the third-order of the external magnetic field:
\begin{align}
    m \propto \frac{h^xh^yh^z}{K^2}.
    \label{Majorana_gap_field_3rd}
\end{align}
We obtain the relation \eqref{Majorana_gap_field_3rd} by regarding the Zeeman energy $-\sum_j \bm h \cdot \bm S_j$ as a perturbation to the Kitaev(-Heisenberg) model.
The third-order perturbation gives rise to the mass term,
\begin{align}
    -\tilde h \sum_{\{i,j,k\}} S_i^xS_j^yS_k^z,
    \label{Majorana_mass_term}
\end{align}
with $\tilde h \propto m \propto h^xh^yh^z/K^2$.
In the presence of the perturbative $\Gamma'$ interaction and the Zeeman energy, 
we obtain the mass term \eqref{Majorana_mass_term} in the second-order process of the perturbation expansion instead of the third-order one without the electric field.
For example, let us consider the M$_1$--M$_2$ bond on the $xy$ plane.
Taking the Zeeman energy $-h^z \sum_j S_j^z$ and the $\Gamma'$ interaction $\Gamma' \sum_{\braket{i,j}_z}(S_i^xS_j^y+S_i^yS_j^x)$ on that bond as perturbations to the Kitaev-Heisenberg Hamiltonian, we obtain~\cite{takikawa_kitaev_off-diagonal_2019,takikawa_kitaev_off-diagonal_2020}
\begin{align}
    \tilde h \propto h^z \Gamma'.
\end{align}
Suppose that we have no $\Gamma'$ interactions in the first place (i.e., $\Gamma'=0$ for $\bm{E}=\bm{0}$).
Then, the Majorana mass $\propto|h^xE^z|$ is the second order about the external electromagnetic fields.
Furthermore, the $\bm{E}$-induced $\Gamma'$ interaction can induce topological phase transitions between gapped quantum spin liquids with different Chern numbers~\cite{takikawa_kitaev_off-diagonal_2020}.
In addition to the $\Gamma'$ intearction, the electric field can also induce the DMI as we saw in this section though we need nonequivalence between ligand sites.
A recent study~\cite{ralko_kitaev_DM_2020} pointed out that the DMI can also drive topological quantum phase transitions between gapped quantum spin liquids under magnetic fields.
Therefore, the electric field turns out to be capable of inducing various topological quantum phase transitions in combination with the magnetic field.

For $\alpha$-RuCl$_3$, we can use parameters $U_d=2.5~\mathrm{eV}$, $U_p=1.5~\mathrm{eV}$, $\Delta_{dp}=5.5~\mathrm{eV}$, $J_{\mathrm{H}}=0.7~\mathrm{eV}$, and $t=0.88~\mathrm{eV}$~\cite{Suzuki2021_rucl3}.
These parameters give $J_{\mathrm{F}}=-3~\mathrm{meV}$.
As we mentioned, our model gives antiferromagnetic $K>0$ but can make it ferromagnetic $K<0$ with slight modifications of the model.
If we ignore the $\bm{E}$ dependence of $J_{\mathrm{F}}$ as we did thus far in this paper, we find that a DC electric field $1$--$10~\mathrm{MV/cm}$ change $|D_{\mathrm{F}}^x|$ and $|\Gamma'|$ only by $10^{-4}$--$10^{-3}$~\% of $J_{\mathrm{F}}$, which is too much underestimated.
Note that the $1$--$10$~MV/cm DC electric field can reduce $J_{\mathrm{F}}$ by $1$--$10$~\%~\cite{furuya_superex}. 
This reduction enhances the change in the ratios $|D_{\mathrm{F}}^x/J_{\mathrm{F}}|$ and $|\Gamma'/J_{\mathrm{F}}|$.
We can further enhance the DC electric-field effect, by including Rashba-SOC-driven hoppings besides the hitherto considered intra-atomic SOC.
1-10~\% change of the SOC will have large impact on the $\bm E$-induced DMI.
This estimation within our perturbation scheme looks experimentally challenging.
However, considering the fact that many multiferroic compounds show magnetoelectric effects~\cite{Tokura2014_review}, we might be optimistic about a chance to obtain larger changes in $|\Gamma'/J_{\rm F}|$ and $|D_{\rm F}^x/J_{\rm F}|$, depending on microscopic details of compounds not taken into account in this paper.

To conclude this section, we stress the fact that the DC electric field yields the terms such as the DMI and the $\Gamma'$ interaction that were forbidden from symmetry in the absence of the electric field.
This emergence of the interaction is possible because the electric field lowers the spatial symmetry.

\section{Rashba spin-orbit coupling}\label{sec:Rashba}

\subsection{Introduction}\label{sec:rashba_hopping}

We now move on to the inter-atomic SOC.
The Rashba SOC will be one of the most famous forms of such SOC.
We can include the Rashba SOC into the theoretical framework through the single-electron Hamiltonian [see Eq.~\eqref{H_single}].
We can effectively regard that the single-electron Hamiltonian $H_1$ contains the Rashba SOC, namely
\begin{align}
    -\alpha_{\mathrm{R}}(\bm{\sigma}\times\bm{p})\cdot\bm{e}(\bm{r}),
    \label{Rashba_SOC}
\end{align}
where $\bm{p}$ is the momentum of the electron in the crystal and $\bm{e}(\bm{r})=\bm{E}(\bm{r})/|\bm{E}(\bm{r})|$ is the unit vector parallel to the electric field $\bm{E}(\bm{r})$.
We can deem $\bm{E}(\bm{r})$ the external electric field, a surface electric field for thin-film materials~\cite{Min_Rashba_2006,Min_Rashba_2006,Caviglia_Rashba_2010}, or an interfacial electric field for field-effect transistors~\cite{Bisri2017,ueno_iwasa_edlt_jpsj}.
The coefficient $\alpha_{\rm R}$ is proportional to the strength of the local electric field:
\begin{align}
    \alpha_{\rm R} \propto |\bm E(\bm r)|.
\end{align}
The Rashba SOC affects the single-electron Hamiltonian and gives spin-flipping hoppings and eventually into the DMI.

\subsection{Example}

\begin{figure}[t!]
    \centering
    \includegraphics[bb = 0 0 800 300, width=\linewidth]{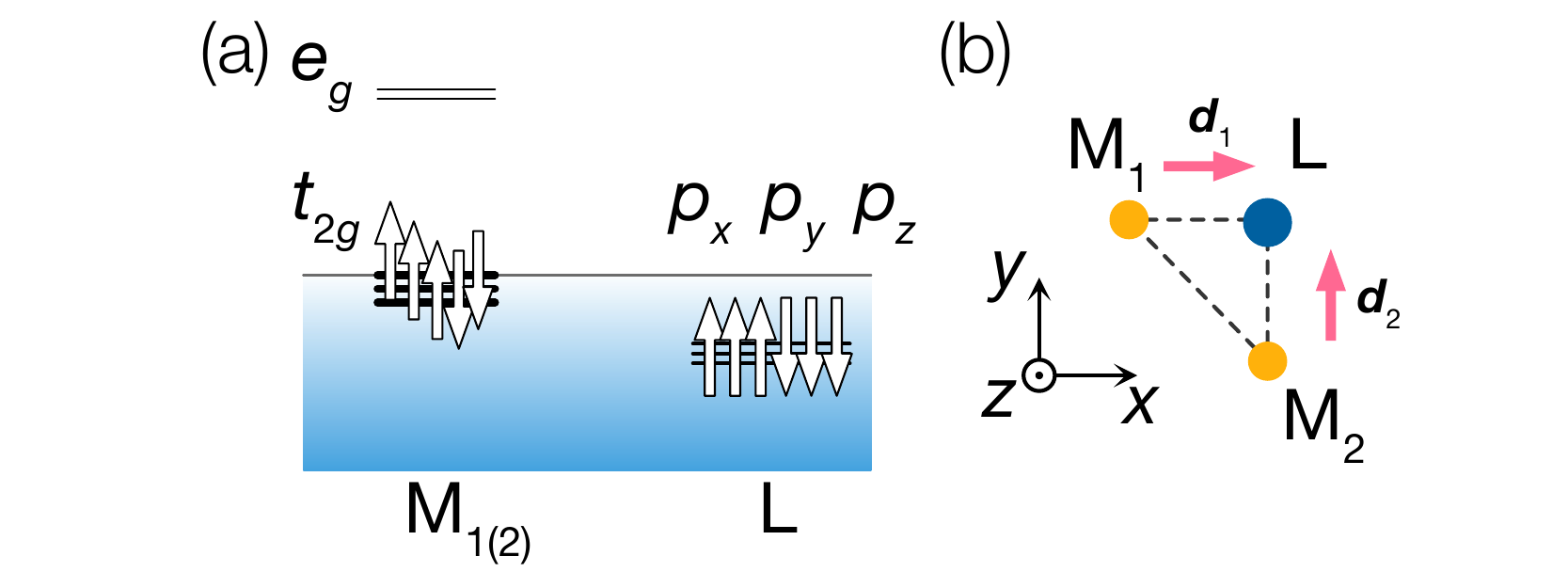}
    \caption{(a) $d^5$ electron configuration of $d$ orbitals under octahedral crystal electric field and $p$ orbitals.
    (b) Spatial configuration of magnetic ions M$_{1,2}$ and ligand ion L.
    The red arrows depict the unit vectors $\bm d_j$ for $j=1,2$ in Eq.~\eqref{rashba_matrix_element}.
    We emphasize that the intra-atomic SOC is not considered here in contrast to the setup of Fig.~\ref{fig:gamma7_config}. 
    We include the inter-atomic Rashba SOC \eqref{Rashba_SOC}.
    }
    \label{fig:rashba_setup_suppl}
\end{figure}

For demonstration, we again consider the isosceles right triangle of Fig.~\ref{fig:rashba_setup_suppl}~(b).
We also inherit the electron configuration, the low-spin $d^5$ configuration under the octahedral crystal electric field [Fig.~\ref{fig:rashba_setup_suppl}~(a)].
The only difference from the previous $J_{\mathrm{eff}}=1/2$ model lies in the SOC mechanism.
In this section, we consider Rashba-SOC-driven hoppings but ignore the $d$-orbital splitting due to the intra-atomic SOC.

The Rashba SOC enters into the single-electron Hamiltonian ${H}_1$:
\begin{align}
    {H}_1 &= \frac{\bm p^2}{2m} + V(\bm r) -\alpha_{\mathrm{R}} \bm e(\bm r) \cdot \bm \sigma \times \bm p,
\end{align}
where $m$ is the mass of the electron and $V(\bm r)$ is the potential that the electron feels.
The single-electron Hamiltonian keeps the $C_{2v}$ symmetry~\cite{kamimura_book} in the absence of the electric and magnetic fields (i.e., $\bm E(\bm r)=\bm h=\bm 0$).
The last term of $H_1$ is nothing but the Rashba SOC.
The Rashba SOC enters into the Hubbard-like model via the hopping amplitude, the matrix element of the single-electron Hamiltonian.
The hopping amplitude between the $p_x$ orbital at L and the $d_a$ orbital at M$_j$ is given by
\begin{align}
    \braket{p_{x,s}|H_1|d_{j,a,s'}}.
\end{align}
When $\bm E(\bm r)=\bm 0$, the hopping amplitude in our three-site model on the isosceles right triangle is reduced to
\begin{align}
    \braket{p_{x,s}|H_1|d_{j,a,s'}} = t \delta_{a,xy}\delta_{j,2} \delta_{s,s'},
    \label{hopping_px_suppl}
\end{align}
with a real constant $t$.
The Rashba SOC adds another term to the right hand side of Eq.~\eqref{hopping_px_suppl}.
For example, $\bm E(\bm r)=E^z\bm e_z$ gives
\begin{align}
    &\braket{p_{x,s}|(- \alpha_{\mathrm{R}}\bm e(\bm r) \cdot \bm \sigma \times \bm p)|d_{j,a,s'}}
    \notag \\
    &= -\alpha_{\mathrm{R}}|\bm k|e^z (\bm \sigma^{ss'}\times (-i\bm d_j))^z \delta_{j,2},
    \label{rashba_matrix_element}
\end{align}
where we replaced the momentum $\bm p=\bm k$ by $-i|\bm k|\bm d_j$ with the unit vector $\bm d_j$ parallel to the M$_j$--L bond [Fig.~\ref{fig:rashba_setup_suppl}~(b)].
$\bm k$ is the wavevector (since we set $\hbar =1$).
The Kronecker's delta $\delta_{j,2}$ appears because the vector $\bm d_j$ is proportional to $\bm e_x$ for $j=1$ and to $\bm e_y$ for $j=2$.

Considering an application to thin-film systems on the $z=0$ plane, we apply an out-of-plane electric field $\bm{E}=E^z\bm{e}_z$ along the $z$ direction~\cite{KaneMele_2005a,KaneMele_2005b}.
The Rashba SOC due to the out-of-plane electric field gives
\begin{align}
    \delta\mathcal{H}_t(\bm{E})&=\sum_{s,s'}[i\lambda p_{y,s}^\dag(\bm{\sigma}^{ss'}\times\bm{d}_1)^zd_{1,xy,s'}
    \notag \\
    &\qquad +i\lambda p_{x,\sigma}^\dag (\bm{\sigma}^{ss'}\times\bm{d}_2)^zd_{2,xy,s'}+\mathrm{H.c.}],
    \label{dHt_Rashba}
\end{align}
with a unit vector $\bm{d}_j$ pointing toward the ligand site from M$_j$.
The hopping amplitude $\lambda$ equals to $\alpha_{\mathrm{R}}|\bm{k}|$.
The DC electric field $\bm E$ adds this spin-dependent hopping \eqref{dHt_Rashba} to the intrinsic spin-independent hopping,
\begin{align}
    \mathcal H_t(\bm 0) &= t\sum_\sigma (p_{y,\sigma}^\dag d_{1,xy,\sigma} + p_{x,\sigma}^\dag d_{2,xy,\sigma} + \mathrm{H.c.}).
\end{align}
The spin Hamiltonian \eqref{H_spin} follows the diagram of Fig.~\ref{fig:diagrams}.
We obtain the $O(|\bm E|)$ correction to the spin Hamiltonian by replacing one of the four $\mathcal H_t(\bm E)$ by $\delta \mathcal H_t(\bm E)$ and the others by $\mathcal H_t(\bm 0)$.
Namely, we find
\begin{widetext}
\begin{align}
    &P \mathcal H_t(\bm E) \biggl(\frac{1}{E_g-\mathcal H_U} Q \mathcal H_t(\bm E) \biggr)^3 P
    \notag \\
    &\approx P \mathcal H_t(\bm 0) \biggl(\frac{1}{E_g-\mathcal H_U} Q \mathcal H_t(\bm 0) \biggr)^3 P + \biggl[P \delta \mathcal H_t(\bm E) \biggl(\frac{1}{E_g-\mathcal H_U} Q \mathcal H_t(\bm 0) \biggr)^3 P + \mathrm{H.c.}\biggr]
    \notag \\
    &\qquad + \biggl[ P \mathcal H_t(\bm 0)\frac{1}{E_g-\mathcal H_U} Q \delta \mathcal H_t(\bm E) \biggl(\frac{1}{E_g-\mathcal H_U} Q\mathcal H_t(\bm 0)\biggr)^2 P 
    +\mathrm{H.c.}
    \biggr].
    \label{H_spin_OE_correction}
\end{align}
\end{widetext}
The spin Hamiltonian for $\bm E=\bm 0$ contains neither magnetically anisotropic terms nor inversion-asymmetric terms because the Hamiltonian $\mathcal H_t(\bm 0) +\mathcal H_U$ for $\bm E=\bm 0$ is magnetically isotropic and inversion symmetric.
The Rashba interaction \eqref{Rashba_SOC} breaks those spin and spatial symmetries at the same time.

Let us show results of the fourth-order perturbation expansion and give detailed derivations in Appendix~\ref{app:rashba}.
The superexchange interaction is again ferromagnetic: $\mathcal{H}_{\mathrm{spin}}(\bm{0})=J_{\mathrm{R}}\bm{S}_1\cdot\bm{S}_2$
with~\cite{furuya_superex}
\begin{align}
    J_{\mathrm{R}}
    &=-8t^2\frac{J_{\mathrm{H}}}{4(U_d-U_p+\Delta_{dp})^2-{J_{\mathrm{H}}}^2}\biggl(\frac{1}{U_d-U_p+\Delta_{dp}}\biggr)^2.
\end{align}

The $\bm{E}$-induced hoppings \eqref{dHt_Rashba} yield the in-plane DMI:
\begin{align}
    \mathcal H_{\rm spin}(\bm E) = J_{\rm R} \bm S_1 \cdot \bm S_2 + \bm{D}_{\mathrm{R}}\cdot\bm{S}_1\times\bm{S}_2
\end{align}
with 
\begin{align}
    D_{\mathrm{R}}^x
    &=32\lambda t^3\biggl(\frac{1}{U_d-U_p+\Delta_{dp}}\biggr)^2\frac{J_{\mathrm{H}}}{4(U_d-U_p+\Delta_{dp})^2-{J_{\mathrm{H}}}^2}, \\
    D_{\mathrm{R}}^y
    &=-32\lambda t^3\biggl(\frac{1}{U_d-U_p+\Delta_{dp}}\biggr)^2\frac{U_d-U_p+\Delta_{dp}}{4(U_d-U_p+\Delta_{dp})^2-{J_{\mathrm{H}}}^2}, \\
    D_{\mathrm{R}}^z
    &=0.
\end{align}
Note that the DMI violates the inversion symmetry about the $x=y$ plane because the Rashba SOC $-\alpha_{\mathrm{R}}(\bm{\sigma}\times\bm{k})^z$ does.

For a parameter set $U_d=3~\mathrm{eV}$, $U_p=2~\mathrm{eV}$, $\Delta_{dp}=5~\mathrm{eV}$, $J_{\mathrm{H}}=1~\mathrm{eV}$, $t=0.1~\mathrm{eV}$, and $\lambda=0.05~\mathrm{eV}$, we obtain $|D_{\mathrm{R}}^x/J_{\mathrm{R}}|\approx0.46$.
We used a value $\lambda=0.05~\mathrm{eV}$, much smaller than $\lambda=\alpha_{\mathrm{R}}|\bm{k}|\propto|\bm{E}(\bm{r})|$ estimated in some materials~\cite{ast2007_rashba,shanavas2014_rashba_tuning}.
Because $\alpha_{\mathrm{R}}$ is well controllable with the electric field~\cite{shanavas2014_rashba_tuning},
we may expect that the external DC electric field can control the DMI substantially both for Kitaev spin liquids and topological spin textures.

\begin{figure}[t!]
    \centering
    \includegraphics[bb = 0 0 1000 850,width=\linewidth]{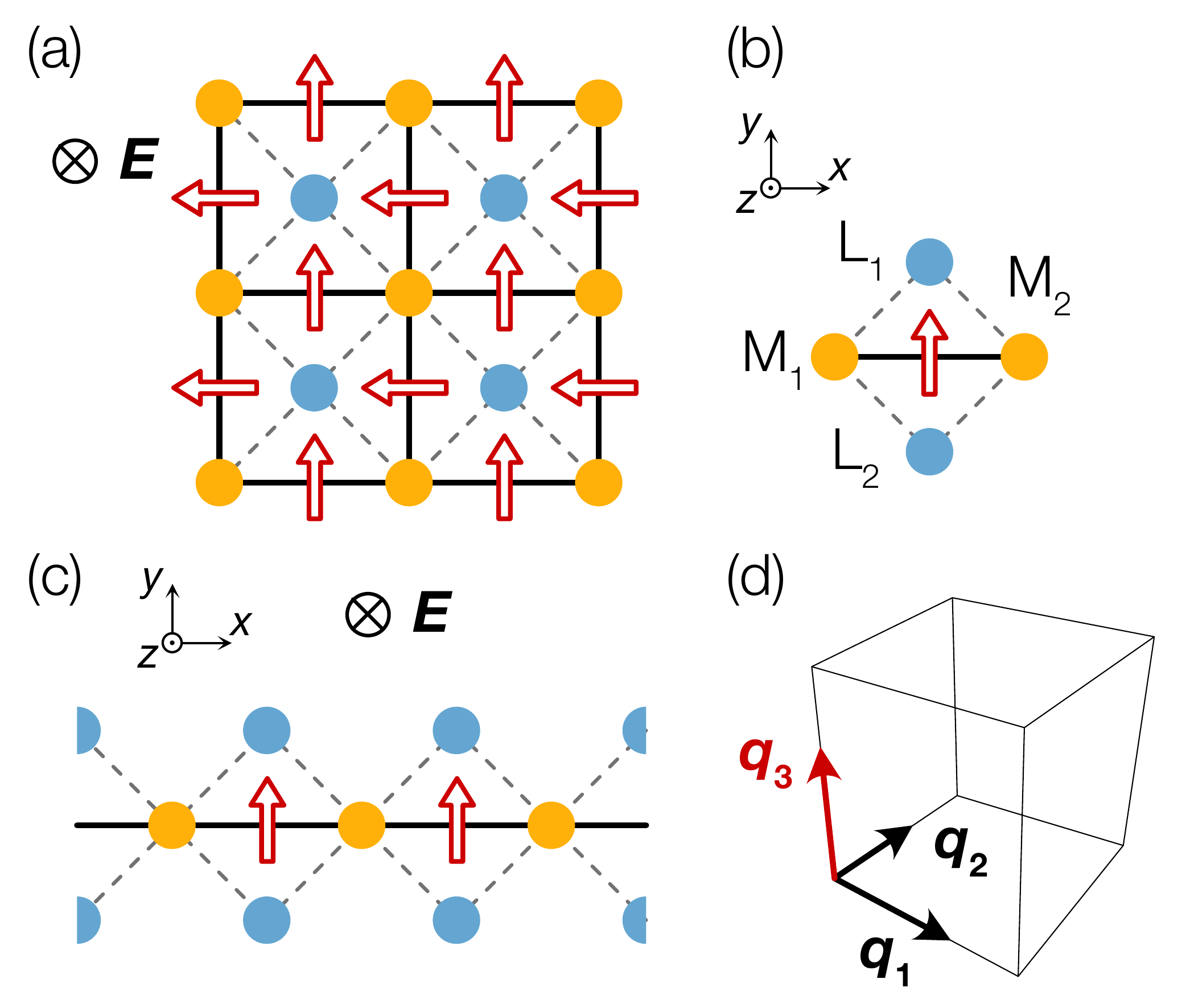}
    \caption{(a) Edge-sharing octahedra ML$_4$ projected to $xy$ plane. Magnetic ions form the square lattice (black solid lines). Intrinsic and $\bm{E}$-induced hoppings occur on dashed gray bonds. 
    (b) Four-site model for nearest-neighbor superexchange interaction of magnetic ions.
    (c) One-dimensional version of square-lattice model (a), ferromagnetic spin chain with uniform DMI.
    (d) $\bm{q}$ vectors of 3$\bm{q}$-hedgehog state~\cite{Fujishiro2019_hedgehog,fujishiro_skyrmion_rev_2020,shimizu_moire}.
    }
    \label{fig:skx_csl}
\end{figure}

\subsection{Strength of external and internal electric fields}\label{sec:estimation}

Let us estimate the required electric-field strength.
Mainly, there are two resources of the DC electric field.
One is to apply it externally using, for example, field-effect transistors.
The other is generated internally by crystal structures.
We call the former external electric fields and the latter internal ones.

Currently, we can realize the external DC field of the strength $\sim 10$~MV/cm using, for example, double-layer transistors~\cite{Bisri2017,ueno_iwasa_edlt_jpsj}.
The internal DC electric field can be even stronger.
Let us consider the octahedral crystal electric field.
The strength of the crystal electric field is typically $\sim 1$~eV~\cite{Suzuki2021_rucl3}.
If the ligand is $\sim 0.1$--$1$~nm away from the magnetic ion, the internal crystal electric field is $\sim 10$--$100$~MV/cm.
This internal crystal electric field is responsible for the Rashba SOC generated on the interface of different materials.
This is the reason why the Rashba SOC can be strong, as we briefly saw in the previous subsection.
We can find another interesting situation in scanning tunneling microscopes (STM)~\cite{Chen2007_STM}.
A tip of the STM induces a DC electric field strong enough to induce the tunneling electron current on the surface.
If the STM tip is $\sim1$~nm distant from the surface and the $\sim1$~V voltage is applied, the surface feels the $\sim10$~MV/cm electric field~\cite{Magtoto2000_stm,Hsu2016}.

A single-cycle THz laser pulse~\cite{hirori_laser_2011,mukai_laser_2016,nicoletti_laser_2016,Fulop2019, sato_laser_dm,sato_floquet_book} will also be useful to generate strong \textit{DC} electric fields. 
Though the laser pulse is not exactly static, the laser pulse can effectively be deemed a DC electric field if the time scale of the electron dynamics ($\sim 10-100$ [fs]) is much faster than the duration of the applied THz laser pulse ($\sim 1$ [ps]).
The strength of the laser pulse will also reach $\sim 1~\mathrm{MV/cm}$~\cite{hirori_laser_2011,mukai_laser_2016,nicoletti_laser_2016,Fulop2019}.
Here we need to be careful about the time dependence of the magnetic anisotropy induced by the THz laser pulse. The THz laser pulse can be regarded as a DC electric field when the electron hoppings are concerned, as we mentioned above. 
Still, since the time scale of the spin dynamics is much slower than that of electron ones, the resultant magnetic anisotropy in the spin Hamiltonian generally shows the time dependence.
The spin dynamics associated with the time-dependent spin Hamiltonian is, in principle, observable by optical measurements such as pump-probe experiments~\cite{kirilyuk2010_ultrafast_rmp}.

\subsection{Applications}\label{sec:appl_rashba}

\subsubsection{Magnetic skyrmion lattice}\label{sec:skyrmion}

We can directly apply our Rashba SOC argument to controls of magnetic skyrmion lattice~\cite{Muhlbauer2009,Yu2010,Seki2012,mochizuki_skx_2012,seki2016skyrmions}.
We exemplify this application by considering a two-dimensional (2D) array of edge-sharing octahedra whose centers have magnetic ions and vertices have ligand ions [Fig.~\ref{fig:skx_csl}~(a)].
The magnetic ions form the square lattice on the $xy$ plane.
We can build this many-body model with the small square plaquette of Fig.~\ref{fig:skx_csl}~(b) made of two isosceles right triangles $\triangle$M$_1$L$_1$M$_2$ and $\triangle$M$_1$L$_2$M$_2$.
Assuming the Rashba SOC \eqref{dHt_Rashba} on the M$_j$--L$_{j'}$ bonds, we obtain the spin Hamiltonian for the building block of Fig.~\ref{fig:skx_csl}~(b):
$\mathcal{H}_{\mathrm{spin}} = J_{\mathrm{R}}\bm{S}_1\cdot\bm{S}_2+D_\perp\bm{e}_\perp\cdot (\bm{S}_1\times \bm S_2)-h_z\sum_jS_j^z$,
where we applied both the electric and magnetic fields parallel to the $z$ axis and $\bm e_\perp = (\bm e_x+\bm e_y)/\sqrt{2}$ is the unit vector perpendicular to the M$_1$--M$_2$ bond.
The vector $D_\perp\bm{e}_\perp$ with $D_\perp=\sqrt{2}(D_{\mathrm{R}}^x+D_{\mathrm{R}}^y)$ is depicted as the red-fringe arrow in Figs.~\ref{fig:skx_csl}~(a), (b), and (c).
With many square plaquettes of Fig.~\ref{fig:skx_csl}~(b), we can build a square-lattice ferromagnet [Fig.~\ref{fig:skx_csl}~(a)],
\begin{align}
    \mathcal H_{\mathrm{SkX}}
    &= -|J_{\mathrm{R}}|\sum_{\bm{r}} (\bm S_{\bm r} \cdot \bm S_{\bm r+\bm e_x} + \bm S_{\bm r} \cdot \bm S_{\bm r+\bm e_y}) - h^z \sum_{\bm r} S_{\bm r}^z
    \notag \\
    &+ D_\perp(E^z) \sum_{\bm r} (\bm S_{\bm r} \times \bm S_{\bm r+\bm e_x}\cdot \bm e_y -\bm S_{\bm r} \times \bm S_{\bm r+\bm e_y}\cdot \bm e_x ).
    \label{H_skx_neel}
\end{align}
The model \eqref{H_skx_neel} realizes the N\'eel-type magnetic skyrmion lattice because a 90$^\circ$ rotation $(S_{\bm{r}}^x,\,S_{\bm{r}}^y,\,S_{\bm{r}}^z)\to(-S_{\bm{r}}^y,\,S_{\bm{r}}^x,\,S_{\bm{r}}^z)$ turns the model \eqref{H_skx_neel} into the well-known model that exhibits the Bloch-type skyrmion lattice~\cite{mochizuki_skx_2012}.
The out-of-plane electric field thus controls the ratio $D_\perp(E^z)/J_{\mathrm{R}}(E^z)$ and allows us to create and annihilate the skyrmion.
Figure~\ref{fig:e_vs_h}~(a) shows the phase diagram of the model \eqref{H_skx_neel}.

\subsubsection{Chiral soliton lattice}\label{sec:CSL}

Building the ML$_2$ chain [Fig.~\ref{fig:skx_csl}~(c)] with the same square plaquette, we can construct a one-dimensional version of the model \eqref{H_skx_neel} with the following Hamiltonian:
\begin{align}
    \mathcal H_{\rm CSL}
    &= -|J_{\rm R}|\sum_j \bm S_j \cdot \bm S_{j+1} - h^z \sum_j S_j^z
    \notag \\
    &\qquad +D_\perp(E^z)\sum_j (\bm e_x \cdot \bm S_j \times \bm S_{j+1}).
    \label{H_CSL}
\end{align}
The spiral order is induced by the ferromagnetic superexchange interaction and the uniform DMI along the spin chain in the $x$ direction.
The transverse magnetic field $h^z$ perpendicular to the DMI makes the ground state the chiral soliton lattice (CSL)~\cite{Kishine2015_csl}.
This ferromagnetic chain exhibits the CSL phase if it is ferromagnetically coupled to proximate spin chains in a three-dimensional crystal~\cite{Togawa2016_csl}.
The ML$_2$ chain is typically realized for $\mathrm{(M,L)=(Cu,O)}$.
The electric field controls the ratio $D_\perp(E^z)/J_{\mathrm{R}}(E^z)$ and turns the ferromagnetic state into the CSL [Fig.~\ref{fig:e_vs_h}~(b)]~\cite{togawa_csl,Kishine2015_csl,Togawa2016_csl}.

\subsubsection{Magnetic hedgehog lattice}\label{sec:hedgehog}

The Rashba SOC is appropriate to drive the chiral structure in a certain direction by generating a uniform DMI.
This feature of the Rashba SOC will also be useful for multiple-$\bm{q}$ states [Fig.~\ref{fig:topological_spin_textures}~(c)]~\cite{Fujishiro2019_hedgehog,fujishiro_skyrmion_rev_2020,shimizu_moire}.
Provided that a magnetic material shows a double-$\bm{q}$ state with the two $\bm{q}$ vectors lying on the $q_x$-$q_y$ plane [$\bm{q}_1$ and $\bm{q}_2$ of Fig.~\ref{fig:skx_csl}~(d)] and the electric field $\bm{E}$ adds another $\bm{q}$ vector ($\bm{q}_3$).
This feature of the electric field will drive the magnetic hedgehog lattice where the triple- or quadruple-$\bm{q}$ spin texture is required.

We can consider another interesting situation, where the single-$\bm q$ state is initially realized without the electric field.
Suppose the single-$\bm q$ configuration along the $z$ axis.
The electric field can add a double-$\bm q$ spin texture on the $xy$ plane to the single-$\bm q$ state, leading to the triple-$\bm q$ state~\cite{nakagawara}.

\section{Summary and Discussions}\label{sec:discussions}

This paper provides the general theoretical foundation to DC electric-field controls of magnetic Mott insulators~\cite{Ohno2000,Romming2013, Matsukura2015,Chen2015,Matsuno2016_DM, Hsu2016,Jiang2018,Huang2018,Chen2019,Xu2020}.
Our formalism treats the electric-field effect on electron orbitals of magnetic and nonmagnetic ions perturbatively.
As Fig.~\ref{fig:diagrams} shows, we considered the fourth-order perturbation expansion about the hopping amplitude because we explicitly take into account the nonmagnetic ligand ion.
The explicit inclusion of nonmagnetic ions is essential to incorporate the DC electric-field effect.
In fact, the electric field lowers the spatial symmetry and induces electron hoppings between magnetic and nonmagnetic ions that were forbidden from the crystalline symmetry.
The symmetry restricts the possible form of the spin Hamiltonian. For example, the DMI is allowed in the presence of the DC electric field even when it is forbidden in the absence of the electric field. The DM vector of such a DMI is obviously dependent on the electric field and approximately proportional to the electric field when the field is perturbative. This phenomenon of the electric-field-induced DMI itself has been known for years.
More specifically, the Rashba-SOC-induced DMI has been discussed for years~\cite{shanavas2014_rashba_tuning,Yu2016_SOC_skyermion_electric,Huang2021_DMI_electric,kim2013_DM_interface,Srivastava2018_trilayer}.
What we did thus far in this paper is to derive the DMI on the microscopic basis.
Note that the symmetry argument implies the induction of the magnetic anisotropy such as the DMI by the electric field but does not tell how large the induced anisotropy is.
We did not rely on the symmetry argument in the derivation of the magnetic anisotropy.
The microscopic derivation is important because it tells us how large the induced mangnetic anisotropy is and what parameters we should tune to obtain more efficient or drastic electric-field effects.

We considered two important cases: the Kitaev material and the magnetic Mott insulator with the Rashba SOC.
In the Kitaev material, we proposed that the electric field generates the DMI and the off-diagonal $\Gamma'$ interaction that can potentially induce phase transitions [e.g., Eqs.~\eqref{H_spin_kitaev_out-of-plane} and \eqref{H_KHgamma'}].
We also showed that the electric-field-induced Rashba SOC gives rise to the DMI, essential to realize topological spin textures such as the magnetic skyrmion, the CSL, and the magnetic hedgehog [e.g., Eqs.~\eqref{H_skx_neel} and \eqref{H_CSL}].
Here, we did not discuss the intra-atomic SOC in the $p$ orbital of the ligand ion because it will be much smaller than the two SOC dealt with in this paper.

For experiments, we propose to use (quasi-)2D systems.
We can generally apply strong enough electric fields to quasi-2D materials 
with state-of-the-art techniques such as double-layer transistors~\cite{ueno_iwasa_edlt_jpsj} and STM~\cite{chen2021_stm_book}. 
The above-mentioned $\alpha$-RuCl$_3$~\cite{Plumb2014_alpha_RuCl3} and other Kitaev-candidate materials such as $\mathrm{Na_2IrO_3}$~\cite{jackeli_kitaev_2009,trebst_kitaev_review_2017} have a quasi-2D layered honeycomb structure.
Quasi-2D materials are compatible with the strong DC electric-field source~\cite{ueno_iwasa_edlt_jpsj,Bisri2017,Romming2013,Hsu2016}.
Besides, the quasi-2D structure allows us to use the strong electric field on the interface to another material.
Generation of strong surface electric fields were already experimentally available~\cite{Bisri2017,ueno_iwasa_edlt_jpsj}.
Reference.~\cite{Matsuno2016_DM} gives an experimental controlling method of the DMI with the interface electric field in the SrRuO$_3$--SrIrO$_3$ bilayer system.
We point out the possible relevance of our study to
2D van der Waals magnets~\cite{Burch2018_vdW_review,Mounet2018_vdW,
motome2023_vdW} since
the 2D van der Waals magnets exhibit large electric-field effects.
Electric-field switching between ferromagnetic and antiferromagnetic states were already experimentally reported~\cite{Jiang2018_vdW,Huang2018_vdW}.
We hope that our paper will stimulate electric-field controls of magnetic \textit{aniostropies} in those quasi-2D magnets.

\section*{Acknowledgments}
The authors are grateful to Tetsuo Hanaguri and Minoru Kanega for fruitful discussions.
S.C.F. and M.S. are supported by a Grant-in-Aid for Scientific Research on Innovative Areas ”Quantum Liquid Crystals” Grant No. JP19H05825. 
S.C.F. is also supported by JSPS Grants-in-Aid for Transformative Research Areas (A) “Extreme Universe,” (Grants Nos. JP21H05191 and
21H05182) and by JSPS KAKENHI (Grants Nos. JP20K03769 and JP21K03465).
M.S. is also supported by JSPS KAKENHI (Grant Nos. 20H01830 and 20H01849) and by a Grant-in-Aid for Scientific Research on Innovative Areas “Evolution of Chiral Materials Science using Helical Light Fields” (Grants No. JP22H05131 and No. JP23H04576) from JSPS of Japan.

\appendix

\section{Field-induced magnetic anisotropies in Kitaev-Heisenberg model}
\label{app:minimal}

\subsection{Effective mapping from $t_{2g}$ orbitals to $J_{\rm eff}=1/2$ doublet}\label{app:zero-field_kitaev}

The $J_{\mathrm{eff}}=1/2$ doublet at M$_j$ are a spin-orbit-entangled superposition of $t_{2g}$ orbitals.
The up state ($\ket{+}_j$) and the down state ($\ket{-}_j$) state of the (pseudo)spin at M$_j$ are given by~\cite{knb,kim_jeff_1/2_2008,Matsuura2013_JJ,Matsuura2014_compass}
\begin{align}
    \ket{+}_j&= \frac{1}{\sqrt 3} (\ket{d_{j,xy,\uparrow}} + \ket{d_{j,yz,\downarrow}} +i\ket{d_{j,zx,\downarrow}}),
    \label{+_Gamma7+}
    \\
    \ket{-}_j &= \frac{1}{\sqrt 3} (\ket{d_{j,xy,\downarrow}} - \ket{d_{j,yz,\uparrow}} +i\ket{d_{j,zx,\uparrow}}),
    \label{-_Gamma7+}
\end{align}
where $\ket{d_{j,a,\sigma}}=d_{j,a,\sigma}^\dag\ket{0}$ is a spin-$\sigma$ state of the $d_a$-orbital electron at M$_j$ and $\ket{0}$ is the vacuum of the creation operator $d_{j,a,\sigma}^\dag$.
This doublet is labeled as $\Gamma_{7+}$ using the Bethe's notation of the double group~\cite{onodera_group_1966}.
Hereafter, we denote $\sigma=\uparrow, \downarrow$ as $\sigma=+,-$, respectively.
Note that under the strong spin-orbit coupling (SOC), we should adopt the so-called $JJ$ coupling scheme instead of the $LS$ coupling one~\cite{Matsuura2013_JJ,Matsuura2014_compass}.
Here, $\bm J$ is a superposition of the angular momentum $\bm L$ and the spin $\bm S$ [see Eq.~\eqref{J_LS}].
Whereas the SOC enters into wave functions of hybridized orbitals in the $LS$-coupling scheme, it does not in the strongly spin-orbit-entangled states \eqref{+_Gamma7+} and \eqref{-_Gamma7+} in the $JJ$-coupling scheme.
This difference comes out of a fact that the former deals with the SOC perturbatively but the latter does nonperturbatively.
In Eqs.~\eqref{+_Gamma7+} and \eqref{-_Gamma7+} formulated within the $JJ$-coupling scheme, the coefficients are fully determined by the crystalline symmetry.

The creation and annihilation operators of the $t_{2g}$-orbital electrons are related to each other in the $J_{\mathrm{eff}}=1/2$ model because the $J_{\mathrm{eff}}=1/2$ doublet is their superposition.
Here, we note that the orbital angular momentum $\bm L=-\bm l_d$ and the spin,
\begin{align}
    \bm S_d :=\frac{\hbar}2 \sum_{a=xy,yz,zx} \sum_{s,s'=\pm} d_{j,a,s}^\dag \bm\sigma^{ss'} d_{j,a,s'},
    \label{Sd_suppl}
\end{align}
form the effective total angular momentum $\bm J_{\mathrm{eff}}$ as follows~\cite{Takayama_review_spin-orbit_2021}.
\begin{align}
    \bm J_{\mathrm{eff}} = - \bm l_d + \bm S_d.
    \label{J_LS}
\end{align}
Hereafter, we set $\hbar=1$ for simplicity.
$\bm S_d$ and $\bm L$ of the $\Gamma_{7+}$ doublet are antiparallel to each other because electrons in the $d^5$ configuration feel the strong SOC $\xi \bm L \cdot \bm S_d$ with large positive $\xi$~\cite{Takayama_review_spin-orbit_2021}.
We can easily confirm
\begin{align}
    J_{\rm eff}^z \ket{\pm}_j &= \pm \frac 12 \ket{\pm}_j.
\end{align}
Since there are few possibilities of confusions, we simply represent this pseudospin $\bm J_{\rm eff}$ as $\bm S_j$ and call it a spin, as we did in the main text.
The spin operator $\bm S_j$ is thus defined as
\begin{align}
    S_j^z &= \frac 12 (\ket{+}_{jj}\bra{+} - \ket{-}_{jj}\bra{-}), \\
    S_j^\pm &= \ket{\pm}_{jj}\bra{\mp}.
\end{align}

Our purpose in this subsection is to write the spin Hamiltonian in terms of this spin operator $\bm S_j$.
For this purpose, we represent creation and annihilation operators of the $t_{2g}$ orbitals in terms of the $J_{\mathrm{eff}}=1/2$ doublet, $d_{j,\Gamma_{7+},\sigma}^\dag$ and $d_{j,\Gamma_{7+},\sigma}$.
They are defined as
\begin{align}
    \ket{+}_j=:d_{j,\Gamma_{7+},+}^\dag\ket{0} , \qquad \ket{-}_j=:d_{j,\Gamma_{7+},-}^\dag\ket{0}.
\end{align}
An  operator, $n_{j,\Gamma_{7+},\sigma} = d_{j,\Gamma_{7+},\sigma}^\dag d_{j,\Gamma_{7+},\sigma}$, counts the number of electrons with the $\sigma$ spin in the $\Gamma_{7+}$ doublet.
We can relate $t_{2g}$-orbital operators $d_{j,a,\sigma}$ with $a=xy,yz,zx$ to the $\Gamma_{7+}$-orbital operator $d_{j,\Gamma_{7+},\sigma}$ as follows.
For example, the operator $d_{j,xy,\sigma}$ satisfies
\begin{align}
	d_{j,xy,+}\ket{\sigma}_j &= \frac 1{\sqrt 3} \delta_{\sigma,+}\ket{0}, \quad d_{j,xy,-}\ket{\sigma}_j = \frac 1{\sqrt 3}\delta_{\sigma,-}\ket{0},
\end{align}
where $\delta_{a,b}$ is Kronecker's delta.
These relations indicate
\begin{align}
	d_{j,xy,\sigma}P  &=  Qd_{j,xy,\sigma}P = Q\biggl(\frac 1{\sqrt 3}  d_{j,\sigma}\biggr)P,
\end{align}
at low energies, where $P=\bigotimes_{j=1,2}\bigl(\ket{+}_{jj}\bra{+}+\ket{-}_{jj}\bra{-}\bigr)$ is the projection operator to the Hilbert subspace spanned by the $\Gamma_{7+}$ doublets and $Q=1-P$ is the projection to its complementary.
Note a simple relation $Pd_{j,a,\sigma}P = 0$ for $a=xy,yz,zx$.
Likewise, we obtain
\begin{align}
	 d_{j,yz,\sigma}P  &= Q d_{j,yz,\sigma}P =  Q\biggl(- \frac 1{\sqrt 3} \sigma  d_{j,-\sigma}\biggr)P, \\
	 d_{j,zx,\sigma}P  &= Q d_{j,zx,\sigma}P= Q \biggl(i \frac 1{\sqrt 3} d_{j,-\sigma}\biggr)P.
\end{align}
These relations lead to
\begin{align}
    \mathcal{H}_t(\bm{0})P
    &= \biggl\{ \frac{t}{\sqrt 3} \sum_{\sigma=\pm} [(p_{y,\sigma}^\dag + i p_{z,-\sigma}^\dag) d_{1,\Gamma_{7+},\sigma} 
    \notag \\
    &\qquad + (p_{x,\sigma}^\dag +\sigma p_{y,-\sigma}^\dag) d_{2,\Gamma_{7+},\sigma}+\mathrm{H.c.}
    ]
    \biggr\}P, \\
    P\mathcal{H}_t(\bm{0})
    &=P \biggl\{ \frac{t}{\sqrt 3} \sum_{\sigma=\pm} [(p_{y,\sigma}^\dag + i p_{z,-\sigma}^\dag) d_{1,\Gamma_{7+},\sigma} 
    \notag \\
    &\qquad + (p_{x,\sigma}^\dag +\sigma p_{y,-\sigma}^\dag) d_{2,\Gamma_{7+},\sigma}+\mathrm{H.c.}
    ]
    \biggr\}.
\end{align}
At low energies, we can abbreviate these relations as Eq.~\eqref{Ht_Kitaev-Heisenberg} in the main text.

\subsection{Evaluation of $I$}\label{app:in-plane_kitaev}

To evaluate the constant $I$, we adopt the wave functions $\braket{\bm r|d_{j,xy,\sigma}}$ and $\braket{\bm r|p_{xy,\sigma}}$ of the hydrogen-like atom in terms of the the polar coordinate $\bm r=r(\sin\theta \cos\phi,\, \sin\theta \sin\phi,\, \cos\theta)$:
\begin{align}
    \braket{\bm r|d_{j,xy,\sigma}}
    &\approx R_{nd}(r)Y_{xy}(\theta,\phi), \\
    \braket{\bm r|p_{x,\sigma}}
    &\approx R_{2p}(r)Y_{x}(\theta,\phi), \\
    R_{3d}(r)
    &= \frac{4}{81\sqrt{30}} \biggl(\frac{Z_M}{a_0}\biggr)^{3/2} \biggl(\frac{Z_M}{a_0}r \biggr)^2 \exp\biggl( - \frac{Z_M}{3a_0}r \biggr), \\
    R_{4d}(r) &= \frac{1}{768\sqrt{5}}\biggl(\frac{Z_M}{a_0}\biggr)^{3/2} \biggl(12 -\frac{Z_Mr}{a_0} \biggr) \biggl(\frac{Z_Mr}{a_0}\biggr)^2
    \notag \\
    &\qquad \times \exp\biggl(-\frac{Z_Mr}{4a_0}\biggr), \\
    R_{2p}(r)
    &= \frac{1}{2\sqrt 6} \biggl(\frac{Z_L}{a_0}\biggr)^{3/2} \frac{Z_L}{a_0}r  \exp\biggl( - \frac{Z_L}{2a_0}r \biggr), \\
    Y_{xy}(\theta,\phi)
    &= \sqrt{\frac{15}{16\pi}} \sin^2\theta \sin(2\phi), \\
    Y_x(\theta,\phi)
    &= \sqrt{\frac{3}{4\pi}} \sin \theta \cos \phi.
\end{align}
Here, $a_0$, $Z_M$, and $Z_L$ are the Bohr radius and the atomic numbers of M$_j$ and L, respectively~\cite{knb,kamimura_book}.
$n=3,4,\cdots$ represent the principal quantum number.
For $n=3$, these wave functions lead to 
\begin{align}
    I 
    &\approx ea_0 \frac{16}{27}Z_M^{7/2}Z_L^{5/2} \biggl(\frac{Z_M}{3}+\frac{Z_L}{2}\biggr)^{-7}.
\end{align}
For $n=4$, it becomes
\begin{align}
    I &\approx ea_0 \frac{\sqrt{3}}{16} Z_M^{7/2}Z_L^{5/2} \biggl(\frac{Z_M}{4}+\frac{Z_L}{2}\biggr)^{-8}.
\end{align}

\section{Rashba spin-orbit interaction in tight-binding models}\label{app:rashba}

\begin{widetext}
To show that the Rashba SOC leads to the DMI, we consider the second term,
\begin{align}
    P \delta \mathcal H_t(\bm E) \biggl(\frac{1}{E_g-\mathcal H_U} Q \mathcal H_t(\bm 0) \biggr)^3 P + \mathrm{H.c.},
\end{align}
of Eq.~\eqref{H_spin_OE_correction}
The second line of Eq.~\eqref{H_spin_OE_correction} is similarly calculated.
The fourth-order perturbation processes are divided into two classes, the processes (a) and (b) of FIG. 18 in Ref.~\cite{furuya_superex}.
The process (a) gives
\begin{align}
    &\biggl[P\delta \mathcal H_t(\bm E)\biggl(\frac{1}{E_g-\mathcal H_U}Q \mathcal H_t(\bm 0) \biggr)^3 P + \mathrm{H.c.}\biggr]_{\text{process (a)}}
    \notag \\
    &= P \biggl[\frac{i\lambda t}{U_d-U_p+\Delta_{dp}} \frac{t^2}{2(U_d-U_p)+2\Delta_{dp}-J_{\mathrm{H}}} \frac{2}{U_d-U_p+\Delta_{dp}}
    \notag \\
    &\quad \cdot \biggl\{\sum_{\sigma_1,\sigma_2,\cdots, \sigma_5=\pm} \frac 12 \delta_{\sigma_1,-\sigma_2}(-\delta_{\sigma_4,\sigma_2}\delta_{\sigma_3,\sigma_1} - \delta_{\sigma_4,-\sigma_2}\delta_{\sigma_3,-\sigma_1}) (\bm \sigma_{\sigma_4\sigma_5}\times \bm d_2)^z d_{2,\sigma_5}d_{1,\sigma_3}d_{2,\sigma_2}^\dag d_{1,\sigma_1}^\dag 
    \notag \\
    &\qquad + \sum_{\sigma_1,\sigma_2,\cdots,\sigma_5} \frac 12 \delta_{\sigma_1,-\sigma_2}(\delta_{\sigma_4,\sigma_2}\delta_{\sigma_3,\sigma_1}+\delta_{\sigma_4,-\sigma_2}\delta_{\sigma_3,-\sigma_1}) (\bm \sigma_{\sigma_4\sigma_5}\times \bm d_1)^z d_{1,\sigma_5}d_{2,\sigma_3} d_{2,\sigma_2}^\dag d_{1,\sigma_1}^\dag
    \biggr\}
    \notag \\
    &\quad + 
    \frac{i\lambda t}{U_d-U_p+\Delta_{dp}} \frac{t^2}{2(U_d-U_p)+2\Delta_{dp}+J_{\mathrm{H}}} \frac{2}{U_d-U_p+\Delta_{dp}}
    \notag \\
    &\quad \cdot \biggl\{ \sum_{\sigma_1,\sigma_2,\cdots,\sigma_5=\pm} \frac 12 \delta_{\sigma_1,-\sigma_2} (-\delta_{\sigma_4,\sigma_2}\delta_{\sigma_3,\sigma_1} + \delta_{\sigma_4,-\sigma_2}\delta_{\sigma_3,-\sigma_1}) (\bm \sigma_{\sigma_4\sigma_5}\times \bm d_2)^z d_{2,\sigma_5} d_{1,\sigma_3} d_{2,\sigma_2}^\dag d_{1,\sigma_1}^\dag 
    \notag \\
    &\qquad + \sum_{\sigma_1,\sigma_2,\cdots,\sigma_5=\pm} \frac 12 \delta_{\sigma_1,-\sigma_2} (\delta_{\sigma_4,\sigma_2}\delta_{\sigma_3,\sigma_1} - \delta_{\sigma_4,-\sigma_2}\delta_{\sigma_3,-\sigma_1}) (\bm \sigma_{\sigma_4\sigma_5}\times \bm d_1)^z d_{1,\sigma_5} d_{2,\sigma_3} d_{2,\sigma_2}^\dag d_{1,\sigma_1}^\dag
    \biggr\}
    \biggr]P,
\end{align}
where we already removed the $p$-orbital operators by using the projection $P$ [see, for example,  Eq.~(A3) of Ref.~\cite{furuya_superex}].
Rewriting the $d$-orbital creation and annihilation operators in terms of the spin $\bm S_j$ ($j=1,2$), we obtain
\begin{align}
    &\biggl[P\delta \mathcal H_t(\bm E)\biggl(\frac{1}{E_g-\mathcal H_U}Q \mathcal H_t(\bm 0) \biggr)^3 P + \mathrm{H.c.}\biggr]_{\text{process (a)}}
    \notag \\
    &= P \biggl[ \frac{-4\lambda t^3}{2(U_d-U_p)+2\Delta_{dp}-J_{\rm H}} \biggl(\frac{1}{U_d-U_p+\Delta_{dp}} \biggr)^2
    \notag \\
    &\qquad 
    \cdot [d_2^y (-S_1^zS_2^y-S_1^yS_2^z) + d_2^x (-S_1^zS_2^x-S_1^xS_2^z)
    +d_1^y(-S_1^zS_2^y-S_1^yS_2^z)+d_1^x(-S_1^zS_2^x-S_1^xS_2^z)]
    \notag \\
    &\qquad +\frac{-4\lambda t^3}{U_d-U_p+2\Delta_{dp}+J_{\rm H}} \biggl(\frac{1}{U_d-U_p+\Delta_{dp}} \biggr)^2
    \notag \\
    &\qquad \cdot [d_2^y(-S_1^zS_2^y+S_1^yS_2^z)+d_2^x(-S_1^zS_2^x+S_1^xS_2^z) + d_1^y(-S_1^zS_2^y+S_1^yS_2^z) + d_1^x(-S_1^zS_2^x+S_1^xS_2^z)]
    \biggr]P.
\end{align}
The process (b) of FIG. 18 in Ref.~\cite{furuya_superex} gives
\begin{align}
    &\biggl[P \delta \mathcal H_t(\bm E) \biggl(\frac{1}{E_g-\mathcal H_U} Q \mathcal H_t(\bm 0) \biggr)^3 P + \mathrm{H.c.}
    \biggr]_{\text{process (b)}}
    \notag \\
    &= P \biggl[\frac{i\lambda t}{U_d-U_p+\Delta_{dp}}\frac{t^2 }{2(U_d-U_p)+2\Delta_{dp}-J_{\rm H}} \frac{2}{U_d-U_p+\Delta_{dp}}
    \notag \\
    &\qquad \cdot \sum_{\sigma_1,\cdots,\sigma_5} \delta_{\sigma_1,\sigma_2} (-\delta_{\sigma_4,\sigma_2}\delta_{\sigma_3,\sigma_1}) (\bm \sigma_{\sigma_4\sigma_5} \times \bm d_2)^z d_{2,\sigma_5} d_{1,\sigma_3} d_{2,\sigma_2}^\dag d_{1,\sigma_1}^\dag
    \notag \\
    &\qquad +\frac{i\lambda t}{U_d-U_p+\Delta_{dp}} \frac{t^2}{2(U_d-U_p)+2\Delta_{dp}-J_{\rm H}} \frac{2}{U_d-U_p+\Delta_{dp}}
    \notag \\
    &\qquad \cdot \sum_{\sigma_1,\cdots,\sigma_5} \delta_{\sigma_1,\sigma_2} (\delta_{\sigma_4,\sigma_2}\delta_{\sigma_3,\sigma_1}) (\bm \sigma_{\sigma_4\sigma_5}\times \bm d_1)^z d_{1,\sigma_5} d_{2,\sigma_3} d_{2,\sigma_2}^\dag d d_{1,\sigma_1}^\dag
    \notag \\
    &\qquad + \mathrm{H.c.}
    \biggr]P
    \notag \\
    &= P \biggl[ \frac{-8\lambda t^3}{2(U_d-U_p)+2\Delta_{dp} -J_{\rm H}} \biggl(\frac{1}{U_d-U_p+\Delta_{dp}}\biggr)^2 (d_2^y S_1^zS_2^y + d_2^x S_1^zS_2^x)
    \notag \\
    &\qquad +\frac{-8\lambda t^3}{2(U_d-U_p)+2\Delta_{dp} - J_{\rm H}} \biggl(\frac{1}{U_d-U_p+\Delta_{dp}}\biggr)^2 (d_1^y S_1^yS_2^z +d_1^x S_1^x S_2^z)
    \biggr]P.
\end{align}
Combining these contributions of the two processes together, we obtain 
\begin{align}
    &P \delta \mathcal H_t(\bm E) \biggl(\frac{1}{E_g-\mathcal H_U} Q \mathcal H_t(\bm 0) \biggr)^3 P + \mathrm{H.c.} 
    \notag \\
    &=P \biggl[\frac{-4\lambda t^3}{2(U_d-U_p)+2\Delta_{dp}-J_{\rm H}} \frac{1}{U_d-U_p+\Delta_{dp}} [(d_1^y-d_2^y)(\bm S_1 \times \bm S_2)^x -(d_1^x-d_2^x)(\bm S_1\times \bm S_2)^y ]
    \notag \\
    &\qquad + \frac{-4\lambda t^3}{2(U_d-U_p)+2\Delta_{dp}+J_{\rm H}} \frac{1}{U_d-U_p+\Delta_{dp}} [(d_1^y+d_2^y)(\bm S_1 \times \bm S_2)^x -(d_1^x+d_2^x)(\bm S_1\times \bm S_2)^y ]
    \biggr]P,
\end{align}
which is nothing but the DMI.
The other terms,
\begin{align}
     P \mathcal H_t(\bm 0)\frac{1}{E_g-\mathcal H_U} Q \delta \mathcal H_t(\bm E) \biggl(\frac{1}{E_g-\mathcal H_U} Q\mathcal H_t(\bm 0)\biggr)^2 P 
    +\mathrm{H.c.}
\end{align}
lead to the same result after repeating a similar procedure to the above one.
Collecting all these contributions, 
we reach the final result:
\begin{align}
    \mathcal H_{\rm spin} &= J \bm S_0 \cdot \bm S_1 + D^x (\bm S_1\times \bm S_2)^x +D^y (\bm S_1\times \bm S_2)^y, \\
    D^x &= -\biggl[(d_1^y-d_2^y) \frac{16\lambda t^3}{2(U_d-U_p)+2\Delta_{dp}-J_{\rm H}} + (d_1^y+d_2^y) \frac{16\lambda t^3}{2(U_d-U_p)+2\Delta_{dp}+J_{\rm H}}\biggr] \biggl(\frac{1}{U_d-U_p+\Delta_{dp}}\biggr)^2, \\
    D^y &= \biggl[(d_1^x-d_2^x) \frac{16\lambda t^3}{2(U_d-U_p)+2\Delta_{dp} -J_{\rm H}} + (d_1^x+d_2^x) \frac{16\lambda t^3}{2(U_d-U_p)+2\Delta_{dp}+J_{\rm H}}
    \biggr]\biggl(\frac{1}{U_d-U_p+\Delta_{dp}}\biggr)^2.
\end{align}
Note that $\bm d_1= \bm e_x$ and $\bm d_2=\bm e_y$ for the spatial configuration of Fig.~\ref{fig:rashba_setup_suppl}~(b).
\end{widetext}

\bibliography{ref.bib}

\end{document}